\documentclass[prb,eqsecnum,preprint]{revtex4}

\usepackage{graphicx}%
\usepackage{dcolumn}
\usepackage{amsmath,amssymb} 

\makeatletter
\def\btt#1{\texttt{\@backslashchar#1}}%
\DeclareRobustCommand\bblash{\btt{\@backslashchar}}%
\makeatother

\begin{document}

\preprint{revised version}

\title[Bipolaron Problem]{Stability of Formation 
of Large Bipolaron: 
Nonrelativistic Quantum Field Theory}
\author{Masao Hirokawa}
\email{hirokawa@math.okayama-u.ac.jp}
\homepage{http://www.math.okayama-u.ac.jp/~hirokawa}
\affiliation{%
The Graduate School of Natural Science and Technology, Okayama University, 
Okayama 700-8530, Japan   \\ 
\qquad \\ 
{\rm Dedicated to Herbert Spohn on the occasion of his 
$60$th birthday}
}%

\date{December 1, 2006}

\begin{abstract}
We are concerned with the stability of formation of 
large bipolaron in a $3$-dimensional ($3$D) crystal. 
This problem is considered in the framework of 
nonrelativistic quantum field theory. 
Thus, the Hamiltonian formalism, 
as Fr\"{o}hlich introduced, 
is employed to describe the bipolaron. 
We approach the problem by characterizing 
some sufficient or necessary conditions 
for the bipolaron being stable. 
This paper gives a full detail of the author's 
talks at ESI, RIMS, and St. Petersburg State Univ. 
in 2005.
\end{abstract}

\maketitle


\section{Introduction}
\label{sec:intro}

The question whether bipolarons can be formed 
has attracted ever a great deal of studies 
since Shafroth\cite{shafroth} in particular showed 
bipolaronic superconductivity takes place when 
the temperature is below that of Bose-Einstein condensation.  
Over the past few decades there have been 
several renewals \cite{anderson,AR,mott,emin} of interest 
in this problem from the point of view 
of high-temperature superconductor.
In this paper, we will focus our mind on the stability of 
formation of large bipolaron.

We will treat two electrons coupled with 
longitudinal optical (LO) phonons in a $3$-dimensional crystal. 
Generally, the electron-phonon interaction dresses 
an electron in a phonon cloud. 
This dressed electron is the so-called polaron. 
We assume the electron-phonon interaction is 
described by the second quantization, 
as Fr\"{o}hlich\cite{FPZ,froehlich} introduced. 
If the Coulomb repulsion between the two electrons 
puts them so far away from each other that 
each electron dresses itself in an individual 
phonon cloud, there is no exchange of phonons 
between the two. 
Namely, there is the only Coulomb repulsion 
between them and thus only two separated single 
polarons are formed in the crystal. 
On the other hand, if the distance 
between the two electrons is so short that the 
phonon-exchange occurs, there is a possibility that 
attraction appears\cite{froehlich0} between them 
and therefore that 
they are bound to each other. 
The bound two polarons is called a bipolaron.

In the light of superconductivity, 
many studies have been performed for 
the large bipolaron.\cite{emin,BGIN,VSPD,VPD,adamowski} 
As for small polaron,\cite{tjablikov,YK,sewell,holstein,LF} 
Alexandrov and Kornilovich\cite{AK} 
clarified the physical properties of the 
small Fr\"{o}hlich polaron. 
Alexandrov and Ranninger\cite{AR} 
showed the possibility 
that small bipolarons might be superconducting. 
Moreover, Alexandrov and Mott,\cite{AM-small} 
and Alexandrov\cite{alexandrov} 
pointed out the mobility of the small bipolaron 
as well as that of the large bipolaron. 
Thus, we are also interested in how the size of 
large bipolaron grows.  

Using the  classical picture described 
in Ref.\onlinecite{IL}, 
we can explain the occurrence of attraction 
between the two electrons in the following. 
Since an electron has negative charge, 
a first electron leaves behind a deformation trail 
in the crystal lattice, 
which affects the positions of the ion cores. 
This trail is associated with an increased density of 
positive charge owing to the ion cores. 
Therefore, it has an attractive effect on a second electron. 
Namely, the lattice deformation causes the attraction 
between the first and second electrons. 
Thus, the problem is whether the attraction is enough 
to bind the two electrons. 

Pekar\cite{pekar} showed that there is no formation 
of the large bipolaron for any optic dielectric constant 
$\epsilon_{\infty}$ and the static dielectric constant 
$\epsilon_{0}$ in $3$D systems.
Also Takada\cite{takada} reached the same result 
in $3$D systems from the weak-coupling 
limit by studying the Bethe-Salpeter equation. 
Obeying the classical theory of phonon 
described in Ref.\onlinecite{AM}, 
we briefly see that two large polarons cannot 
be bound to each other from the point of view 
of binding energy 
$2E_{\mbox{\tiny {\rm SP}}}^{\mbox{\tiny {\rm AM}}}(r) 
- E_{\mbox{\tiny {\rm BP}}}^{\mbox{\tiny {\rm AM}}}(r)$, 
where $E_{\mbox{\tiny {\rm SP}}
}^{\mbox{\tiny {\rm AM}}}(r)$ 
is the total energy of a large (single) polaron 
in a sphere of radius $r > 0$, and 
$E_{\mbox{\tiny {\rm BP}}}^{\mbox{\tiny {\rm AM}}}(r)$ 
the total energy of a large bipolaron in the sphere. 
Because the binding energy is always negative in classical theory. 
More precisely, with the help of expressions 
of $E_{\mbox{\tiny {\rm SP}}}^{\mbox{\tiny {\rm AM}}}(r)$ 
and $E_{\mbox{\tiny {\rm BP}}}^{\mbox{\tiny {\rm AM}}}(r)$ 
given in Ref.\onlinecite{AM}, 
we can easily derive the equation: 
\begin{eqnarray}
- \left(
2E_{\mbox{\tiny {\rm SP}}}^{\mbox{\tiny {\rm AM}}}(r) 
- E_{\mbox{\tiny {\rm BP}}}^{\mbox{\tiny {\rm AM}}}(r)
\right) 
= \sqrt{2}\,\alpha 
\left( 
\frac{1}{1-\eta} - 1\right)
\frac{1}{r}
\label{eq:Intr-BE}
\end{eqnarray}
for $\alpha$ (the coupling constant of electrons 
and the phonon field) and 
$\eta := \epsilon_{\infty}/\epsilon_{0}$ 
(the ionicity of the crystal) 
under the natural units. 
So, we can say that 
the binding energy is negative in the 
classical theory because $0 < \eta < 1$. 
A similar form to RHS of Eq.(\ref{eq:Intr-BE}) 
will play an important role in Secs.\ref{sec:SLWCR}, 
\ref{sec:formation}, and \ref{sec:nonnegative-binding} 
(see Eqs.(\ref{negative-r-binding}), 
(\ref{eq:cBP}), 
(\ref{eq:VG-functional}), 
(\ref{eq:appearance-theta1}), 
and (\ref{eq:appearance-theta2})). 
Thus, the attraction derived in classical theory 
is too weak to bind the two polarons.

On the other hand, 
Vinetskii and Giterman\cite{VG} found that 
the bipolaron is formed only for very small 
ionicities $\eta$. 
Emin\cite{emin} also pointed out 
a possibility of formation of large bipolaron for 
sufficiently large coupling constants $\alpha$ 
in addition to small ionicities $\eta$. 
Our argument using Fr\"{o}hlich's interaction 
in this paper is based on many studies in 
quantum theory by prior 
literature\cite{BGIN,VSPD,VPD,adamowski} 
to derive the effective attraction. 
Bassani {\it et al.}\cite{BGIN} showed the large bipolaron 
formation for large $\alpha$ and small $\eta$ 
by estimating the binding energy. 
Verbist {\it et al.}\cite{VSPD} showed 
the stable region of $\eta$ for the formation of 
the large bipolaron in the strong-coupling limit. 
It is worthy of note that the Feynman path-integral 
approach\cite{VPD,HT} is also useful for the bipolaron problem.

In this paper, to derive an effective attraction 
in the same framework as 
in studies,\cite{BGIN,VSPD,VPD,adamowski} 
we will adopt the notion of 
Feynman's classical virtual 
phonon\cite{feynman} 
instead of the deformation 
trail in the crystal lattice.
In addition, we will take the image of 
Peeters and Devreese's classical 
bipolaron\cite{PD} into our argument. 
In their theory of the classical bipolaron, 
the two electrons rotate in 
a circle around a common fixed center. 
We now denote the radius of the circle 
by $r_{\mbox{\tiny {\rm PD}}}$. 
Then, the classical bipolaron has to be localized 
in the closed ball 
$\overline{B}(r_{\mbox{\tiny {\rm PD}}})$ of 
radius $r_{\mbox{\tiny {\rm PD}}}$ centered 
at the fixed point. 
The radius $r_{\mbox{\tiny {\rm PD}}}$ should 
turn out long when the Coulomb repulsion 
beats the attraction caused by the phonon field. 
Conversely, it should turn out short 
when the attraction wins 
over the Coulomb repulsion. 
So, if the attraction between the two electrons 
is so weak that the Coulomb repulsion makes 
the interior of the ball $\overline{B}(r_{\mbox{\tiny {\rm PD}}})$ 
completely contains the whole crystal, 
then we cannot hope that the classical bipolaron 
is formed in the crystal. 
One of our ideas is to take this image 
into nonrelativistic quantum field theory 
by reorganizing the way we have ever done.\cite{hirokawa-rims} 
Thus, we have to introduce the notion of 
the size of the region in which the two electrons 
constructing bipolaron live. 
For the ground state energy 
$E_{\mbox{\tiny {\rm BP}}}$ 
of bipolaron and 
the ground state energy 
$E_{\mbox{\tiny {\rm SP}}}$ 
of single polaron, 
we introduce an energy $E(r)$ coming from 
the phonon field for every $r > 0$ as: 
\begin{eqnarray}
2E_{\mbox{\tiny SP}} - E_{\mbox{\tiny BP}} 
= 
E(r) - \frac{U}{r}.
\label{eq:emin}
\end{eqnarray}
That is, we express the binding energy 
with the energy $E(r)$ and the Coulomb 
repulsive potential $U/r$. 
Here, $r$ stands for the distance between 
the two electrons now. 
Taking account of Emin's work \cite{emin1} 
and Salje's,\cite{salje} 
$E(r)$ is probably non-negative. 
Thus, main purpose in this paper 
is to estimate $E(r)$ so that the stability 
of formation of large bipolaron can be characterized 
in terms of the estimated $E(r)$ and also $\eta$.

In Sec.\ref{sec:Hamiltonians}, we will introduce 
the Hamiltonian of the bipolaron 
which has the Fr\"{o}hlich interaction.  
In Sec.\ref{sec:BE} we will show a device to 
introduce a parameter $\theta \ge 0$ 
which controls the coupling strength 
between the regimes of the weak- and strong-coupling 
theories. To find this parameter, 
we will develop Lieb and Thomas' method.\cite{LT}
In Sec.\ref{sec:SLWCR} we will argue 
spatial localization\cite{HHS,hirokawa-rims} of 
the relative motion of bipolaron in the weak-coupling regime 
by estimating $E(r)$.  
Especially, we will estimate the distance 
between the two electrons 
in the bipolaron from below 
by adopting Lieb's idea\cite{lieb} 
into nonrelativistic quantum field theory. 
Thus, we will understand how the ionicity raises the 
size of the bipolaron in the weak-coupling regime. 
In Sec.\ref{sec:formation}, combining 
the notion of Feynman's virtual phonon\cite{feynman} 
and the image of Peeters and Devreese's classical 
bipolaron,\cite{PD} 
we will derive two effective Hamiltonians 
in quantum theory from the original one. 
The effective Hamiltonians describe those 
in the strong-coupling regime. 
With the help of these effective Hamiltonians 
we will find a sufficient condition for the bipolaron 
formation in terms of $E(r)$ or $\eta$ 
and give a lower and an upper bounds to the ground state 
energy $E_{\mbox{\tiny {\rm BP}}}$. 
In Sec.\ref{sec:SLSCR} we will argue the spatial 
localization in the strong-coupling regime. 
In Sec.\ref{sec:nonnegative-binding} 
we will consider a sufficient condition 
for the positive binding energy.

\section{Hamiltonians}
\label{sec:Hamiltonians}

The total energy of the bipolaron consisting of 
two electrons coupled with 
the LO phonons is described 
by the Hamiltonian $H_{\mbox{\tiny {\rm BP}}}$:
\begin{eqnarray}
H_{\mbox{\tiny {\rm BP}}} = 
H_{{\mathrm {el}}\mbox{-}{\mathrm {el}}} + 
H_{\mathrm {ph}} 
+ H_{{\mathrm {el}}\mbox{-}{\mathrm {ph}}},   
\label{eq:ourHamiltonian}
\end{eqnarray}
where $H_{{\mathrm {el}}\mbox{-}{\mathrm {el}}}$ 
is the energy of two electrons with 
the Coulomb repulsion between them, 
$H_{\mathrm {ph}}$ the free energy of 
the phonon field, 
and $H_{{\mathrm {el}}\mbox{-}{\mathrm {ph}}}$ 
 Fr\"{o}hlich's interaction\cite{FPZ,froehlich} 
derived through the second quantization 
of the Coulomb long-range interaction: 
\begin{eqnarray}
&{}& 
H_{{\mathrm {el}}\mbox{-}{\mathrm {el}}} 
= \sum_{j=1,2}
\frac{1}{2m}p_{j}^{2}  
+ \frac{U}{|x_{1}-x_{2}|}, 
\label{eq:el-el} \\ 
&{}& 
H_{\mathrm {ph}} 
= 
\sum_{k}\hbar\omega_{k}a^{\dagger}_{k}a_{k},   
\label{eq:ph} \\ 
&{}& 
H_{{\mathrm {el}}\mbox{-}{\mathrm {ph}}} 
= 
\sum_{j=1,2}
\sum_{k}
\bigg\{
V_{k}e^{ik\cdot x_{j}}a_{k} 
+ 
V_{k}^{*}e^{-ik\cdot x_{j}}a^{\dagger}_{k}
\bigg\}. 
\label{eq:Froehlich-interaction}
\end{eqnarray} 
In Eq.(\ref{eq:el-el}), 
the position and momentum operators of the $j$th 
electron ($j = 1,2$) of mass $m$ 
are denoted by $x_{j}$ and $p_{j}$, respectively, 
so $p_{j} = - i\hbar\nabla_{x_{j}} $. 
The strength of the Coulomb repulsion is 
designated by the symbol $U$, so 
$U \equiv e^{2}/\epsilon_{\infty}$
for the electric charge $e$ and 
the optic dielectric constant $\epsilon_{\infty}$.
In Eq.(\ref{eq:ph}), $a_{k}$ and $a_{k}^{\dagger}$ are 
the annihilation and creation operators, respectively, 
of the LO phonon with the momentum $\hbar k$. 
Then $a_{k}$ and $a_{k}^{\dagger}$ satisfy 
the canonical commutation relation, 
$[ a_{k}, a_{\ell}^{\dagger}] = \delta_{k\ell}$, 
because phonons are bosons. 
The LO phonons can be assumed to be dispersionless, 
$\omega_{k} = \omega_{\mbox{\rm {\tiny LO}}}$.  
In Eq.(\ref{eq:Froehlich-interaction}), 
$V_{k}$ is given by $V_{k} := 
-\, i\hbar\omega_{\mbox{\rm {\tiny LO}}}
\left(4\pi\alpha r_{{}_{\mathrm {fp}}}/
k^{2}V \right)^{1/2}$ for the crystal volume $V$ 
and the free polaron radius $r_{{}_{\mathrm {fp}}} 
\equiv \left(\hbar/2m\omega_{\mbox{\rm 
{\tiny LO}}}\right)^{1/2}$. 
The dimensionless electron-phonon 
coupling constant is given by 
\begin{eqnarray}
\alpha := 
\frac{1}{\hbar\omega_{\mbox{\rm {\tiny LO}}}} 
\frac{e^{2}}{2}
\left(
\frac{1}{\epsilon_{\infty}} - 
\frac{1}{\epsilon}_{0}\right)
\frac{1}{r_{{}_{\mathrm {fp}}}}, 
\label{eq:coupling-constant}
\end{eqnarray}
where $\epsilon_{0}$ is the static dielectric constant.

Concerning the Hamiltonian $H_{\mbox{\tiny {\rm BP}}}$ 
of bipolaron, we make some remarks. 
Since the ionicity $\eta$ of 
the crystal is defined by $\eta := \epsilon_{\infty}
/\epsilon_{0}$, it satisfies $0 < \eta < 1$. 
In terms of $\eta$, the strength of the Coulomb repulsion 
is rewritten as $U = \sqrt{2}\,\alpha/(1-\eta)$. 
Since we consider the two-body system of large 
polarons, the wave vector $k$ in $\sum_{k}$ 
runs over the first Brillouin zone.\cite{BB,pines,AM2} 
This fact makes some noticeable differences 
between phonon and photon 
(cf. Table 23.4 of Ref.\onlinecite{AM2}). 
The primitive cell is usually given by 
the first Brillouin zone for a crystal. 
However, it is not always to chose 
the first Brillouin zone for another solid. 
When a solid has metallic properties, 
we have to take the Fermi surface into account. 
In this case, we employ a reduced zone scheme.\cite{AM2}

We use the natural units $\hbar = m 
= \omega_{\mbox{\tiny LO}} = 1$ from now on. 
Using the conversion\cite{kittel} of sums 
to integrals, 
we estimate $\sum_{k}|V_{k}|^{2}$ at 
$\sqrt{2}\,\alpha K/\pi$ as:
\begin{eqnarray}
\sum_{k}|V_{k}|^{2} 
\approx \frac{N}{V_{\mathrm {c}}^{*}}
\int_{|k|\le K}d^{3}k\, 
\frac{4\pi\alpha r_{{}_{\mathrm {fp}}}}{Vk^{2}} 
= \frac{V}{(2\pi)^{3}}
\int_{|k|\le K}d^{3}k\, 
\frac{4\pi\alpha r_{{}_{\mathrm {fp}}}}{Vk^{2}} 
= \frac{\sqrt{2}\,\alpha}{\pi}K, 
\label{eq:(III-0)}
\end{eqnarray}
where $V_{\mathrm {c}}^{*}$ is the volume of the primitive 
cell in the reciprocal lattice and 
$V_{\mathrm {c}}^{*} = (2\pi)^{3}/V_{\mathrm {c}}$ for 
$V_{\mathrm {c}}$, the volume of the primitive cell 
in the direct lattice. 
We denoted by $N$ the number of primitive cells which 
is contained in the crystal volume $V$. 
So, when integral $\int d^{3}k$ is over the first 
Brillouin zone, $K$ means the radius of a sphere 
of the first Brillouin zone. 
For example, in the Debye interpolation scheme\cite{AM2} 
for a harmonic crystal, $K$ is given by $K = k_{\mathrm {D}}$, 
which is defined by 
$k_{\mathrm {D}}^{3} = (3/4\pi) 
(2\pi)^{3}N/V = 6\pi^{2}N/V$. 
This equation says, as is well known, 
there is the relation between 
$k_{\mathrm {D}}$ and $k_{\mathrm {F}}$ 
(the radius of the Fermi surface), 
i.e., 
$k_{\mathrm {D}} = (2/Z)^{1/3}k_{\mathrm {F}}$, 
where $Z$ is the nominal valence. 
When the integral $\int d^{3}k$ can be extended from the 
first Brillouin zone to the whole $k$-space, 
$K$ plays the role of an ultraviolet (UV) cutoff.  

Using the approximation\cite{kittel} of 
the Fourier expansion, 
$V/(4\pi|x|) \approx \sum_{k}e^{ik\cdot x}/k^{2}$, 
we obtain 
\begin{eqnarray}
\sum_{k}|V_{k}|^{2}e^{ik\cdot x} 
\approx  \frac{\alpha}{\sqrt{2}\,|x|}. 
\label{eq:FE}
\end{eqnarray}
We often use this approximation 
(\ref{eq:FE}) in this paper. 

We also use 
the coordinate of the center-of-mass, 
$X_{1} = (x_{1} + x_{2})/2$, 
and the coordinate of the relative motion, 
$X_{2} = x_{1} - x_{2}$, in this paper. 
Each momentum is given by $P_{j} = - i\nabla_{X_{j}}$, 
$j = 1, 2$. 
So, we have $P_{1} = p_{1} + p_{2} = M_{1}\dot{X}_{1}$ 
and $P_{2} = (p_{1} - p_{2})/2 = M_{2}\dot{X}_{2}$, 
respectively, where we set the masses as 
$M_{1} = 2$ and $M_{2} = 1/2$. 
Then, $H_{\mbox{\tiny BP}}$ is unitary-equivalent to 
\begin{eqnarray}
\widetilde{H}_{\mbox{\tiny BP}} 
= 
\widetilde{H}_{{\mathrm {el}}
\mbox{-}{\mathrm {el}}}
+ H_{\mathrm {ph}} 
+ \widetilde{H}_{{\mathrm {el}}\mbox{-}{\mathrm {ph}}},
\label{eq:Hcr}
\end{eqnarray}
where 
\begin{eqnarray*} 
\widetilde{H}_{{\mathrm {el}}
\mbox{-}{\mathrm {el}}}
&=& \sum_{j=1,2}\frac{1}{2M_{j}}P_{j}^{2} 
+ \frac{U}{|X_{2}|}, \\ 
\widetilde{H}_{{\mathrm {el}}\mbox{-}{\mathrm {ph}}}
&=& 
\sum_{k}
c(X_{2},k) \left\{ 
V_{k}e^{ik\cdot X_{1}}a_{k} 
+ 
V_{k}^{*}e^{-ik\cdot X_{1}}a^{\dagger}_{k}
\right\},  
\end{eqnarray*}
with 
$$
c(X_{2},k) 
= 2\cos\frac{k\cdot X_{2}}{2} = 
e^{ik\cdot X_{2}} + e^{-ik\cdot X_{2}}. 
$$

As for $H_{\mbox{\tiny {\rm BP}}}$, 
we note the following. 
The bipolaron Hamiltonian $H_{\mbox{\tiny {\rm BP}}}$ 
and the total 
momentum $\Pi_{\mathrm {tot}} 
= p_{1} + p_{2} + \sum_{k}ka_{k}^{\dagger}
a_{k}$ are commutable, i.e., 
$\left[ \Pi_{\mathrm {tot}} , 
H_{\mbox{\tiny {\rm BP}}}\right] = 0$, 
because of the translation invariance of 
$H_{\mbox{\tiny {\rm BP}}}$. 
We can show that $H_{\mbox{\tiny {\rm BP}}}$ 
has no ground sate in the standard mathematical 
representation because of the continuous symmetry 
of the translation invariance.\cite{HM} 
It is known in general that if a Hamiltonian $H$ has a 
continuous symmetry (that is, $\left[ H, \Pi\right] = 0$ 
for the generator $\Pi$ of a transformation) and moreover 
$H$ has a ground state, 
there are two possibilities: 
the transformation $e^{it\Pi}$ might 
leave the ground state invariant for every real number $t$, 
which is called manifest symmetry, or otherwise, 
it might map a ground state $\Psi_{0}$ to 
another ground state $e^{it\Pi}\Psi_{0}$, which is called 
hidden symmetry. 
In the latter case, we can show that there are 
infinitely degenerate ground states.\cite{HM} 
Moreover, the Nambu-Goldstone theorem 
suggests that the spontaneous symmetry breaking 
would occur and then the Nambu-Goldstone bosons 
would appear. 
We conjecture that they appear as 
acoustic phonons in bipolaron, 
taking account of the result in Ref.\onlinecite{LRU}.

To avoid such a kind of situation coming from 
the continuous symmetry and thus to give 
a possibility that $H_{\mbox{\tiny {\rm BP}}}$ 
has a ground state, we can consider, 
for example, the case where we restrict 
the electrons' movement into the crystal 
and the case where we nail down the center-of-mass 
of the two electrons at a point $Q$. 
For both cases, we only have to employ 
$$
H_{{\mathrm {el}}\mbox{-}{\mathrm {ph}}}(\rho) 
= 
\rho(x_{1}+x_{2})
\sum_{j=1,2}
\sum_{k}
\left\{
V_{k}e^{ik\cdot x_{j}}a_{k} 
+ 
V_{k}^{*}e^{-ik\cdot x_{j}}a^{\dagger}_{k}
\right\}    
$$ 
instead of $H_{{\mathrm {el}}\mbox{-}{\mathrm {ph}}}$ 
in $H_{\mbox{\tiny {\rm BP}}}$, where $\rho(x)$ is 
a function satisfying $0 \le \rho(x) \le 1$. 
Namely, in those cases the total Hamiltonian is 
$H_{\mbox{\tiny {\rm BP}}} =
H_{{\mathrm {el}}\mbox{-}{\mathrm {el}}} + 
H_{\mathrm {ph}} 
+ H_{{\mathrm {el}}\mbox{-}{\mathrm {ph}}}(\rho)$.  
In the former, $\rho(x)$ is defined so that $\rho(x) = 0$ 
outside the crystal and $\rho(x) = 1$ inside the crystal. 
In the latter, 
$\rho(x)$ is defined by 
$\rho(x) \equiv  \rho_{Q}(x)$, 
where $\rho_{Q}(x) := 1$ if $x = 2Q$;\, 
$\rho_{Q}(x) := 0$ if $x \ne 2Q$. 
Then, we can separate the center-of-mass motion from 
$H_{\mbox{\tiny {\rm BP}}}$ and thus obtain its 
relative motion. 
Taking $\rho(x) = 1$ for every $x$ 
restores the Hamiltonian 
$H_{{\mathrm {el}}\mbox{-}{\mathrm {el}}} + 
H_{\mathrm {ph}} 
+ H_{{\mathrm {el}}\mbox{-}{\mathrm {ph}}}(\rho)$ 
to the original $H_{\mbox{\tiny {\rm BP}}}$. 
When we introduce $\rho(x)$ into 
$\widetilde{H}_{\mbox{\tiny BP}}$, 
we employ $\widetilde{H}_{{\mathrm {el}}
\mbox{-}{\mathrm {ph}}}
(\rho)$ instead of 
$\widetilde{H}_{
{\mathrm {el}}\mbox{-}
{\mathrm{ph}}}$, 
where 
$\widetilde{H}_{{\mathrm {el}}\mbox{-}
{\mathrm {ph}}}(\rho)$ is given by 
putting $\rho(2X_{1})$ 
in front of $\sum_{k}$ in the definition of 
$\widetilde{H}_{{\mathrm {el}}
\mbox{-}{\mathrm {ph}}}$. 
When we fix the center-of-mass at $Q$, 
our target Hamiltonian is 
$\widetilde{H}_{\mbox{\tiny {\rm BP}}}^{\mathrm {rel}} 
= \widetilde{H}_{{\mathrm {el}}\mbox{-}{\mathrm {el}}} 
+ H_{\mathrm {ph}} 
+ \widetilde{H}_{{\mathrm {el}}\mbox{-}
{\mathrm {ph}}}(\rho_{Q})$.
As used in Ref.\onlinecite{BGIN}, 
there is another method, which is given by fixing 
the total momentum $\Pi_{\mathrm {tot}}$ 
at a real number in the spectrum 
of $\Pi_{\mathrm {tot}}$. 
For example, in the case where $Q$ is the origin $O$,   
we can obtain the 
Hamiltonian $\widetilde{H}^{\Pi=0}_{\mbox{\tiny {\rm BP}}} 
:= e^{iP_{\mathrm {ph}}\cdot X_{1}}
\widetilde{H}_{\mbox{\tiny {\rm BP}}} 
e^{-iP_{\mathrm {ph}}\cdot X_{1}} 
= 
P_{2}^{2}/2M_{2} 
+ U/|X_{2}| 
+ H_{\mathrm {ph}} 
+ P_{\mathrm {ph}}^{2}/2M_{1} 
+ \widetilde{H}_{{\mathrm {el}}\mbox{-}
{\mathrm {ph}}}(\rho_{O})$, 
where $P_{\mathrm {ph}} 
:= \sum_{k}ka_{k}^{\dagger}a_{k}$.

\section{Weak- and Strong-Coupling Regimes} 
\label{sec:BE}

In this section, a parameter $\theta \ge 0$ 
is introduced to control the strength of 
the coupling constant $\alpha$ in 
the Hamiltonian of single polaron. 
We will use this parameter $\theta$ 
for the Hamiltonian of bipolaron as well as 
that of single polaron throughout this paper.

Let us define $E_{\mbox{\tiny BP}}$, 
the ground state energy of bipolaron, 
as that of $H_{\mbox{\tiny BP}}$, i.e., 
$E_{\mbox{\tiny BP}} := 
\inf{\mathrm {Spec}}(H_{\mbox{\tiny BP}})$. 
We denote the energy spectrum of a Hamiltonian 
$H$ by ${\mathrm {Spec}}(H)$. 
The ground state energy of $H$ always means 
$\inf{\mathrm {Spec}}(H)$ throughout our argument.

We denote the ground state energy of 
single polaron by $E_{\mbox{\tiny SP}}$, 
i.e., $E_{\mbox{\tiny SP}} := 
\inf{\mathrm {Spec}}(H_{\mbox{\tiny SP}}^{(j)})$, 
for the single-polaron Hamiltonian: 
\begin{eqnarray}
H_{\mbox{\tiny SP}}^{(j)} = 
\frac{1}{2}p_{j}^{2} 
+ 
\sum_{k}a^{\dagger}_{k}a_{k} 
+ 
\sum_{k}
\left\{ 
V_{k}e^{ikx_{j}}a_{k} 
+ 
V_{k}^{*}e^{-ikx_{j}}a^{\dagger}_{k}
\right\},  
\qquad j = 1, 2.   
\end{eqnarray}

Through many studies\cite{feynman,LLP,VSPD,DV,LT} 
we know there are two regimes with respect to 
powers of $\alpha$ in $E_{\mbox{\tiny {\rm SP}}}$: 
$E_{\mbox{\tiny SP}} \propto - \alpha$ 
in the weak-coupling limit  
and $E_{\mbox{\tiny SP}} \propto - \alpha^{2}$ 
in the strong-coupling limit. 
When the wave vector $k$ in $\sum_{k}$ 
runs over an infinite lattice, 
Lieb and Thomas\cite{LT} controlled the UV cutoff 
$K$ by $\alpha$ rigorously to give a lower bound to 
$E_{\mbox{\tiny SP}}$ 
in the strong-coupling theory. 
But, since our $K$ is now the radius of the 
first Brillouin zone, we add a device to their idea. 
Using a unitary operator, 
we introduce the preceding parameter 
$\theta \ge 0$ to control the coupling strength 
between the regimes of the weak- and 
the strong-coupling theories. 
Namely, we represent the two regimes 
by $\theta$ in the following. 
We fix $\alpha$ in the weak-coupling theory 
first ($\theta = 0$). 
As we switch on $\theta$ ($\theta > 0$) 
and make it large, we have the electron-phonon 
coupling constant $\alpha_{\theta} := \alpha\theta$ 
in the strong-coupling theory and 
$\alpha_{\theta}$ increases. 
The method proposed here will 
be basically employed to obtain 
effective Hamiltonians 
for the bipolaron 
in Sec.\ref{sec:formation}.

We define a canonical transformation with 
the generator $G(\theta) := i\theta
\sum_{k}\{ V_{k}a_{k} - V_{k}^{*}a_{k}^{\dagger}\}$. 
Then, our unitary-transformed Hamiltonian is  
$H_{\mbox{\tiny {\rm SP}}}^{(j)}(\theta) 
:=  
e^{iG(\theta)}H_{\mbox{\tiny SP}}^{(j)}e^{-iG(\theta)}$. 
We note that $E_{\mbox{\tiny {\rm SP}}}$ is  
the ground state energy of 
$H_{\mbox{\tiny {\rm SP}}}^{(j)}(\theta)$ 
since $H_{\mbox{\tiny {\rm SP}}}^{(j)}(\theta)$ 
is unitary-equivalent to 
$H_{\mbox{\tiny {\rm SP}}}^{(j)}$. 
We show that $E_{\mbox{\tiny {\rm SP}}}$ 
can be estimated in the following. 
It is easy to show that {\it for $\alpha$ in 
the weak-coupling regime}  
\begin{eqnarray}
-\, \frac{\sqrt{2}\,\alpha}{\pi}K 
\le E_{\mbox{\tiny {\rm SP}}} \le 0.
\label{theorem:BE1-3}
\end{eqnarray} 
We can show that $E_{\mbox{\tiny {\rm SP}}} \propto 
-\, \alpha$ for small coupling constants $\alpha$ 
in the weak-coupling regime 
by obeying a perturbative method as well known 
or using a nonperturbative one such as in Theorem 2 
of Ref.\onlinecite{HHS2}. 
Meanwhile, it is shown in this section  that {\it 
for so large $\theta$ 
that $\alpha_{\theta} := 
\alpha\theta$ is in the strong-coupling regime,  
a lower and an upper bounds to
$E_{\mbox{\tiny {\rm SP}}}$ 
are given as: 
\begin{eqnarray}
-\, c_{\mbox{\tiny {\rm SP}}}\alpha_{\theta}^{2} 
-\, \Bigg| 
1 - \frac{1}{\theta}
\Bigg|
\frac{\sqrt{2}\,\alpha_{\theta}}{\pi}K 
\le 
E_{\mbox{\tiny {\rm SP}}} 
\le 
-\, c_{\mbox{\tiny {\rm SP}}}\alpha_{\theta}^{2} 
+ \theta\frac{\sqrt{2}\,\alpha_{\theta}}{\pi}K,   
\label{theorem:BE1-2}
\end{eqnarray} 
with the approximation (\ref{eq:FE}), 
where $c_{\mbox{\tiny {\rm SP}}} 
= 0.108513\cdots$}. 
We note that the estimate (\ref{theorem:BE1-2}) 
in the strong-coupling 
regime tends to the estimate (\ref{theorem:BE1-3}) 
in the weak-coupling one as $\theta$ approaches $0$.  

To show the inequality (\ref{theorem:BE1-2}) 
we need some mathematical arguments. 
We now introduce a mathematical 
parameter $R' > 0$ into $H_{\mbox{\tiny SP}}^{(j)}$ 
so that $\lim_{R'\to\infty}
H_{\mbox{\tiny SP}}^{(j)}(R') 
= H_{\mbox{\tiny SP}}^{(j)}$ (in the norm resolvent 
sense\cite{RS}) as follows: 
\begin{eqnarray} 
H_{\mbox{\tiny SP}}^{(j)}(R') 
:= 
\frac{1}{2}p_{j}^{2} 
+ 
\sum_{k}a^{\dagger}_{k}a_{k} 
+ 
\gamma_{1}(x_{j})
\sum_{k}
\left\{ 
V_{k}e^{ikx_{j}}a_{k} 
+ 
V_{k}^{*}e^{-ikx_{j}}a^{\dagger}_{k}
\right\},  
\qquad j = 1, 2,   
\end{eqnarray}
where $\gamma_{1}(x)$ is 
a smooth real-valued function 
which satisfies $0 \le \gamma_{1}(x) \le 1$ 
and $\gamma_{1}(x) 
= \gamma_{1}(-x)$ for every $x$. 
The parameter $R' > 0$ is introduced 
so that $\gamma_{1}(x) = 1$ for $|x| \le R'/2$; 
$\gamma_{1}(x) = 0$ for $|x| \ge R'$. 
We denote by $E_{\mbox{\tiny SP}}(R')$ 
the ground state energy of $H_{\mbox{\tiny SP}}^{(j)}(R')$, 
i.e., $E_{\mbox{\tiny SP}}(R') := 
\inf{\mathrm {Spec}}(H_{\mbox{\tiny SP}}^{(j)}(R'))$. 
Then, it automatically follows that 
$\lim_{R'\to\infty}E_{\mbox{\tiny SP}}(R') 
= E_{\mbox{\tiny SP}}$. 

In the same way as for $H_{\mbox{\tiny {\rm SP}}}^{(j)}(\theta)$ 
we define 
$H_{\mbox{\tiny {\rm SP}}}^{(j)}(\theta,R')$ 
by $H_{\mbox{\tiny {\rm SP}}}^{(j)}(\theta,R') 
:= e^{iG(\theta)}H_{\mbox{\tiny {\rm SP}}}^{(j)}(R')
e^{-iG(\theta)}$. 
Then, through the approximation (\ref{eq:FE}) 
we approximate $H_{\mbox{\tiny {\rm SP}}}^{(j)}
(\theta,R')$ as:  
\begin{eqnarray*}
\nonumber 
H_{\mbox{\tiny {\rm SP}}}^{(j)}(\theta,R')  
&\approx& 
\frac{1}{2}p_{j}^{2} 
- 2\theta\,\gamma_{1}(x_{j})
\sum_{k}|V_{k}|^{2}e^{ik\cdot x_{j}} \\ 
\nonumber 
&{}& 
+H_{\mathrm {ph}} 
+ 
\sum_{k}\left\{ 
V_{k}\left( \gamma_{1}(x_{j})
e^{ik\cdot x_{j}}
- \theta\right)a_{k} 
+ V_{k}^{*}\left( \gamma_{1}(x_{j})
e^{-ik\cdot x_{j}} 
- \theta\right)
a_{k}^{\dagger}
\right\} \\ 
&{}&  
+ \theta^{2}\sum_{k}|V_{k}|^{2}. 
\end{eqnarray*}
We estimate $E_{\mbox{\tiny {\rm SP}}}(R')$ 
from below in the first step and from above 
in the next step. 
After obtaining both estimates, 
taking $R'\to\infty$ yields the estimate 
(\ref{theorem:BE1-2}). 

We arbitrarily fix a normalized phonon state $\Psi$ 
and a normalized wave function $\psi(x_{j})$ 
of the electron satisfying 
$\lim_{|x_{j}|\to\infty}\psi(x_{j}) = 0$. 
The brackets 
$\langle\!\langle\quad\rangle\!\rangle$ 
stands for an averaging over 
the wave function $\psi(x_{j})$. 
We define a unitary operator $U_{j,\alpha\theta}$ 
so that $U_{j,\alpha\theta}^{*}p_{j}U_{j,\alpha\theta} 
= \alpha\theta p_{j}$ and 
$U_{j,\alpha\theta}^{*}x_{j}U_{j,\alpha\theta} 
= x_{j}/\alpha\theta$. 
Then, the term $\langle\!\langle 
U_{j,\alpha\theta}^{*}
H_{\mbox{\tiny SP}}^{(j)}(\theta, R')
U_{j,\alpha\theta}
\rangle\!\rangle$ in the equation,    
$$ 
\Bigl<\left(U_{j,\alpha\theta}\psi
\right)\Psi\Big|\, 
H_{\mbox{\tiny SP}}^{(j)}(\theta,R') 
\, 
\Big|\left( U_{j,\alpha\theta}\psi
\right)\Psi\Bigr> 
= 
\Bigl<\Psi\Big| 
\langle\!\langle 
U_{j,\alpha\theta}^{*}
H_{\mbox{\tiny SP}}^{(j)}(\theta, R')
U_{j,\alpha\theta}
\rangle\!\rangle 
\, \Big|\Psi\Bigr>, 
$$ 
can be approximated as:  
\begin{eqnarray*}
&{}& 
\langle\!\langle 
U_{j,\alpha\theta}^{*}
H_{\mbox{\tiny SP}}^{(j)}(\theta, R')
U_{j,\alpha\theta}
\rangle\!\rangle \\ 
&\approx& 
\frac{\alpha^{2}\theta^{2}}{2}
\langle\!\langle 
p_{j}^{2}
\rangle\!\rangle 
- 2\theta\sum_{k}|V_{k}|^{2}
\langle\!\langle 
\gamma_{1}(x_{j}/\alpha\theta)
e^{ik\cdot x_{j}/\alpha\theta}
\rangle\!\rangle 
+ \theta^{2}\sum_{k}|V_{k}|^{2} \\ 
&{}& 
+ H_{\mathrm {ph}} 
+ \sum_{k}
\Bigl\{
V_{k}
\langle\!\langle 
\gamma_{1}(x_{j}/\alpha\theta)
e^{ik\cdot x_{j}/\alpha\theta} 
- \theta 
\rangle\!\rangle 
a_{k} 
+  
V_{k}^{*}
\langle\!\langle 
\gamma_{1}(x_{j}/\alpha\theta)
e^{- ik\cdot x_{j}/\alpha\theta} 
- \theta 
\rangle\!\rangle 
a_{k}^{\dagger}
\Bigr\}. 
\end{eqnarray*}
Thus, it is estimated from below as: 
\begin{eqnarray*}
\langle\!\langle 
U_{j,\alpha\theta}^{*}
H_{\mbox{\tiny SP}}^{(j)}(\theta, R')
U_{j,\alpha\theta}
\rangle\!\rangle  
&\ge& 
\frac{\alpha^{2}\theta^{2}}{2}
\langle\!\langle 
p_{j}^{2}
\rangle\!\rangle 
- 
\theta\sum_{k}|V_{k}|^{2}
\big| 
\langle\!\langle 
\gamma_{1}(x_{j}/\alpha\theta)
e^{ik\cdot x_{j}/\alpha\theta}  
\rangle\!\rangle 
\big|^{2} \\ 
&{}&\qquad 
+ 
(\theta - 1)\sum_{k}|V_{k}|^{2}
\big| 
\langle\!\langle 
\gamma_{1}(x_{j}/\alpha\theta)
e^{ik\cdot x_{j}/\alpha\theta}  
\rangle\!\rangle 
\big|^{2}. 
\end{eqnarray*}
Since $|\langle\!\langle 
\gamma_{1}(x_{j}/\alpha\theta)
e^{ik\cdot x_{j}/\alpha\theta}  
\rangle\!\rangle| \le 1$, 
we arrive at the inequality: 
\begin{eqnarray}
\nonumber 
&{}& 
\bigl<\left(U_{j,\alpha\theta}\psi
\right)\Psi\big|\, 
H_{\mbox{\tiny SP}}^{(j)}(\theta,R')\, 
\big|\left( U_{j,\alpha\theta}\psi
\right)\Psi\bigr> \\ 
\nonumber 
&\ge&  
\frac{\alpha^{2}\theta^{2}}{2}
\int d^{3}x\, 
\big|\nabla_{x}\psi(x)\big|^{2} \\ 
\nonumber 
&{}& 
- 
\theta\sum_{k}|V_{k}|^{2}
\int\int d^{3}xd^{3}y\, 
\gamma_{1}(x/\alpha\theta)
\gamma_{1}(y/\alpha\theta)
|\psi(x)|^{2}
|\psi(y)|^{2}
e^{ik\cdot (x-y)/\alpha\theta}  \\ 
&{}&\qquad 
- |\theta - 1|\sum_{k}|V_{k}|^{2}.
\label{eq:BE-UPB1}
\end{eqnarray}
Approximation (\ref{eq:FE}) shows 
\begin{eqnarray}
\nonumber 
&{}& 
\sum_{k}|V_{k}|^{2}
\int\int d^{3}xd^{3}y\, 
\gamma_{1}(x/\alpha\theta)\gamma_{1}(y/\alpha\theta)
\big|\psi(x)\big|^{2}
\big|\psi(y)\big|^{2}
e^{ik\cdot (x - y)/\alpha\theta} \\ 
&\le& 
\frac{\alpha^{2}\theta}{\sqrt{2}}
\int\int d^{3}xd^{3}y\, 
\frac{\big|\psi(x)\big|^{2}
\big|\psi(y)\big|^{2}}{|x-y|}, 
\label{eq:BE-UPB2}
\end{eqnarray}
where inequalities, 
$0 \le \gamma_{1}(x/\alpha\theta), 
\gamma_{1}(y/\alpha\theta) \le 1$, 
were used in the above.
Thus, the inequalities, 
(\ref{eq:BE-UPB1}) and (\ref{eq:BE-UPB2}), 
lead us to the estimate from below:   
\begin{eqnarray*} 
\bigl<\left(U_{j,\alpha\theta}\psi\right)
\Psi\big|\, 
H_{\mbox{\tiny SP}}^{(j)}(\theta,R')\, 
\big|\left( U_{j,\alpha\theta}\psi
\right)\Psi\bigr> 
\ge 
\alpha^{2}\theta^{2}{\mathcal E}_{\mbox{\tiny {\rm P}}}(\psi) 
- |\theta - 1|\sum_{k}|V_{k}|^{2}, 
\end{eqnarray*}
where ${\mathcal E}_{\mbox{\tiny {\rm P}}}(\psi)$ is the 
Pekar functional,\cite{miyake1,lieb0,spohn,GL} i.e., 
$$
{\mathcal E}_{\mbox{\tiny {\rm P}}}(\psi) 
:= 
\frac{1}{2}\int d^{3}x\big|\nabla_{x}\psi(x)\big|^{2} 
- \frac{1}{\sqrt{2}} 
\int\int d^{3}xd^{3}y\, 
\frac{\big|\psi(x)\big|^{2}
\big|\psi(y)\big|^{2}}{|x-y|}.
$$
Lieb\cite{lieb0} proved 
that there is a unique and smooth minimizing 
$\psi(x)$ in $c_{\mbox{\tiny {\rm SP}}} 
:= -\, \inf_{\psi, \langle\!\langle\psi\rangle\!\rangle = 1}
{\mathcal E}_{\mbox{\tiny {\rm P}}}(\psi)$ up to translations. 
Then, using the estimate (\ref{eq:(III-0)}) 
and the estimate of $c_{\mbox{\tiny {\rm SP}}}$ shown by 
Miyake\cite{miyake1} and by Gerlach and L\"{o}wen,\cite{GL} 
we obtain 
\begin{eqnarray}
E_{\mbox{\tiny {\rm SP}}}(R') 
\ge 
- c_{\mbox{\tiny {\rm SP}}}\alpha_{\theta}^{2} 
- \Bigg| 1 - 
\frac{1}{\theta}\Bigg|
\frac{\sqrt{2}\,\alpha_{\theta}}{\pi}K,\qquad 
c_{\mbox{\tiny {\rm SP}}} 
= 0.108513\cdots.
\label{eq:BE-UPB3}
\end{eqnarray}
This inequality works for the upper bound in 
the inequality (\ref{theorem:BE1-2}). 

Let us proceed to the proof of 
the upper bound next. 
For the estimate from above, 
we prepare 
\begin{eqnarray*}
&{}& 
H_{\mathrm {eff}}^{(j)}(\theta,R') 
:= \frac{1}{2}p_{j}^{2} 
- 2\theta\gamma_{1}(x_{j})
\sum_{k}|V_{k}|^{2}e^{ik\cdot x_{j}} 
\approx  \frac{1}{2}p_{j}^{2} 
- \frac{\sqrt{2}\,\alpha\theta}{|x_{j}|} 
+ {\mathrm {error}}_{{}_{R'}}(x_{j},\theta), \\ 
&{}& 
{\mathrm {error}}_{{}_{R'}}(x_{j},\theta) 
:= (1 - \gamma_{1}(x_{j}))
\frac{\sqrt{2}\,\alpha\theta}{|x_{j}|},
\end{eqnarray*}
by using the approximation (\ref{eq:FE}).

For $r \ge 0$ we define a function $C_{0}(r)$ by 
\begin{eqnarray}
C_{0}(r) := (1 + r)e^{- r}.
\label{eq:C_{0}}
\end{eqnarray} 
Since $C_{0}(r') < C_{0}(r) < C_{0}(0) = 1$ 
for $0 < r < r'$ and $C_{0}(\infty) = 
\lim_{r\to\infty}C_{0}(r) = 0$,
we take a unique point $r_{*}$ so that 
$C_{0}(r_{*}) = 1 - \sqrt{c_{\mbox{\tiny {\rm SP}}}}$.
We set $C_{1}(\mu, R')$ and $R_{\mu}$ 
as $C_{1}(\mu, R') := C_{0}(\mu R')$ 
and $R_{\mu} := r_{*}/\mu$, respectively. 
Then, we have a simple inequality: 
\begin{eqnarray}
C_{1}(\mu , R') 
\le 
1 - \sqrt{c_{\mbox{\tiny {\rm SP}}}}, 
\label{eq:C_{1}}
\end{eqnarray}
provided $R_{\mu} \le R'$.
Using the wave function 
$\psi_{\mu}(x_{j})$ defined by 
$\psi_{\mu}(x_{j}) := 
\sqrt{\mu^{3}/\pi}
\,\, e^{-\mu|x_{j}|}$ 
for $\mu > 0$ 
as a trial function, 
we can estimate $\big<\psi_{\mu}\big|\, 
H_{\mathrm {eff}}^{(j)}(\theta,R')\, 
\big|\psi_{\mu}\big>$ from above:    
\begin{eqnarray}
\big<\psi_{\mu}\big|\, 
H_{\mathrm {eff}}^{(j)}(\theta,R')\, 
\big|\psi_{\mu}\big>  
\le 
\frac{\mu^{2}}{2} - \sqrt{2}\,\alpha_{\theta}\mu 
+ \sqrt{2}\,\alpha_{\theta}\mu C_{1}(\mu,R'),  
\label{eq:(BE-2)}
\end{eqnarray}
where 
we used $0 \le 1 - \gamma_{1}(x_{j}) \le 1$. 
Evaluating RHS of the inequality 
$E_{\mbox{\tiny {\rm SP}}}(R') 
\le \big<\psi_{\mu}\Omega_{\mathrm {ph}} \big|\, 
H_{\mbox{\tiny {\rm SP}}}^{(j)}(\theta,R')\, 
\big|\psi_{\mu}\Omega_{\mathrm {ph}}\big>$ 
for the phonon vacuum $\Omega_{\mathrm {ph}}$, 
it follows from 
the inequalities (\ref{eq:C_{1}}) and 
(\ref{eq:(BE-2)}) that  
\begin{eqnarray*} 
E_{\mbox{\tiny {\rm SP}}}(R') 
\le 
\frac{1}{2}
\left( \mu - \sqrt{2c_{\mbox{\tiny {\rm SP}}}}\,
\alpha_{\theta}\right)^{2} 
- c_{\mbox{\tiny {\rm SP}}}\alpha_{\theta}^{2} 
+ \theta^{2}\sum_{k}|V_{k}|^{2} 
\end{eqnarray*}
for every $\mu > 0$ and $R' \ge R_{\mu}$. 
Setting the parameter $\mu$ as 
$\mu = \mu_{*} := \sqrt{2c_{\mbox{\tiny {\rm SP}}}}\,
\alpha_{\theta}$ 
and using estimate (\ref{eq:(III-0)}), 
we arrive at the following upper bound:  
\begin{eqnarray} 
E_{\mbox{\tiny {\rm SP}}}(R') 
\le 
\inf_{\mu>0}\big<\psi_{\mu}\Omega_{\mathrm {ph}} 
\big|\, H_{\mbox{\tiny {\rm SP}}}^{(j)}(\theta,R')\, 
\big|\psi_{\mu}\Omega_{\mathrm {ph}}\big>  
\le    
- c_{\mbox{\tiny {\rm SP}}}\alpha_{\theta}^{2} 
+ \theta\frac{\sqrt{2}\,\alpha_{\theta}}{\pi}K. 
\label{eq:(BE-3)}
\end{eqnarray}

Both bounds (\ref{eq:BE-UPB3}) and (\ref{eq:(BE-3)}) 
lead us to the conclusion that 
if $R' \ge R_{\mu_{*}} \equiv 
r_{*}/\mu_{*} = r_{*}/\sqrt{2c_{\mbox{\tiny {\rm SP}}}}\, 
\alpha_{\theta}$ 
and $\theta > 0$ then 
\begin{eqnarray}
-\, c_{\mbox{\tiny {\rm SP}}}\alpha_{\theta}^{2} 
-\, \Bigg| 
1 - \frac{1}{\theta}
\Bigg|
\frac{\sqrt{2}\,\alpha_{\theta}}{\pi}K 
\le 
E_{\mbox{\tiny {\rm SP}}}(R') 
\le 
-\, c_{\mbox{\tiny {\rm SP}}}\alpha_{\theta}^{2} 
+ \theta\frac{\sqrt{2}\,\alpha_{\theta}}{\pi}K. 
\label{theorem:BE1-2'}
\end{eqnarray}
Taking the limit $R'\to \infty$ in the above, 
we reach our desired estimate (\ref{theorem:BE1-2}).

\section{Spatial Localization in Weak-Coupling Regime}
\label{sec:SLWCR}

In this section, we consider the case 
where $\alpha$ is in the weak-coupling regime. 
As noted in Sec.\ref{sec:Hamiltonians}, 
the Hamiltonian formalism in the mathematical 
representation does not show us 
the existence of any ground state in the 
state space because of the translation 
invariance of $H_{\mbox{\tiny {\rm BP}}}$ with 
$\rho(x) \equiv 1$. Thus, we employ 
$H_{{\mathrm {el}}\mbox{-}{\mathrm {ph}}}(\rho)$ 
in $H_{\mbox{\tiny {\rm BP}}}$ so that 
$\rho$ breaks the translation invariance. 
Then, in the weak-coupling regime, 
we seek a necessary condition for 
$H_{\mbox{\tiny {\rm BP}}}$ having a bound state and 
investigate spatial localization 
of bipolaron. 
We note that we can obtain the same results for 
$\widetilde{H}^{\mathrm {rel}}_{\mbox{\tiny {\rm BP}}}$ 
and $\widetilde{H}^{\Pi=0}_{\mbox{\tiny {\rm BP}}}$  
as described below. 
Although we do not prove this remark in this paper, 
it is proved in the same way as shown 
for $H_{\mbox{\tiny {\rm BP}}}$.

In Ref.\onlinecite{hirokawa-rims} we showed that 
if the quantum particle is not spatially localized, 
then the particle dressed in the cloud of bosons 
cannot exist. 
In that problem, divergent soft bosons for 
the infrared catastrophe cause 
the spatial nonlocalization. 
From this point of view, 
we study how the Coulomb repulsion 
between the two electrons arrests 
the spatial localization in bipolaron.

Let us proceed with the definition of spatial localization. 
We say that {\it the relative motion of 
the bipolaron in a bound state $\Psi_{n}$ 
is spatially localized in the closed ball} 
$\overline{B}(r) := \left\{ X_{2}\, |\, |X_{2}| \le r\right\}$ 
if $\big<\Psi_{n}\big|
\Psi_{n}\big>^{-1}
\big<\Psi_{n}\big| 
|x_{1} - x_{2}|
\big|\Psi_{n}\big> 
\le r$.  
In this paper we will use the phrase that {\it 
a bound state $\Psi_{n}$ of the bipolaron 
with  the size less than $r > 0$ does not exist} 
when the relative motion of the bipolaron in $\Psi_{n}$ is not 
spatially localized in $\overline{B}(r)$. 
For a ground state $\Psi_{0}$ 
we define the distance 
$d_{\mbox{\tiny {\rm BP}}}(\Psi_{0})$ 
between the two electrons in bipolaron  by
\begin{eqnarray}
d_{\mbox{\tiny {\rm BP}}}(\Psi_{0}) := 
\big<\Psi_{0}\big|
\Psi_{0}\big>^{-1}
\big<\Psi_{0}\big| 
|x_{1} - x_{2}|
\big|\Psi_{0}\big>. 
\label{eq:bipolaron-radius-0}
\end {eqnarray}

To estimate $E(r)$ in Eq.(\ref{eq:emin}) 
under the weak-coupling regime, 
we define a positive constant $E_{\mathrm {w}}(\alpha)$ by    
\begin{eqnarray}
E_{\mathrm {w}}(\alpha) := 4\sum_{k}|V_{k}|^{2} 
= \frac{4\sqrt{2}\,\alpha}{\pi}K 
= \frac{8\sqrt{2}\,\alpha}{\Lambda}
\label{theorem1}
\end{eqnarray}
for every $\alpha > 0$, 
where $\Lambda$ is a wave length defined by 
$\Lambda := 2\pi/K$. 
It is easy to check $E_{\mbox{\tiny BP}} \le 0$ 
by evaluating RHS of the inequality  
$E_{\mbox{\tiny BP}} \le 
\inf_{\psi}\big<\psi\Omega_{\mathrm {ph}}\big|\, 
\widetilde{H}_{\mbox{\tiny BP}}\, 
\big|\psi\Omega_{\mathrm {ph}}\big>$ 
for normalized wave functions $\psi$ 
of the two electrons and the phonon 
vacuum $\Omega_{\mathrm {ph}}$. 
The simple operator inequality $H_{\mathrm {ph}} 
+ \widetilde{H}_{{\mathrm {el}}\mbox{-}{\mathrm {ph}}} 
\ge  -\, E_{\mathrm {w}}(\alpha)$ shows that  
$H_{\mbox{\tiny {\rm BP}}} 
\ge -\, E_{\mathrm {w}}(\alpha)$, 
making a lower bound to the ground state energy:  
\begin{eqnarray}
E_{\mbox{\tiny {\rm BP}}} \ge -\, E_{\mathrm {w}}(\alpha).
\label{eq:lower-bound-weak}
\end{eqnarray} 
We can show $E_{\mbox{\tiny {\rm BP}}} \propto 
-\, \alpha$ for small $\alpha$ 
in the weak-coupling regime by a perturbative or nonperturbative 
means in the same way as in the case of 
$E_{\mbox{\tiny {\rm SP}}}$. 

We prove in this section 
that {\it if the bipolaron has 
a bound state with eigenenergy $E$ so that 
its relative motion is spatially localized in 
$\overline{B}(r)$, 
then there is the energy inequality:}
\begin{eqnarray}
E_{\mathrm {w}}(\alpha) + E  \ge \frac{U}{r} 
= \frac{\sqrt{2}\,\alpha}{(1-\eta)r}. 
\label{eq:assumption-n}
\end{eqnarray}
Our result says that 
\begin{eqnarray}
d_{\mbox{\tiny {\rm BP}}}(\Psi_{0}) \ge 
d_{\mbox{\tiny {\rm BP}}}^{\mathrm {low}}
:= \frac{U}{E_{\mathrm {w}}(\alpha) 
+ E_{\mbox{\tiny {\rm BP}}}} 
\ge \frac{\Lambda}{8(1-\eta)}
\label{eq:lower-estimate-bipolaron-radius}
\end{eqnarray}
{\it for any ground state $\Psi_{0}$}. 
Thus, we realize that the distance 
$d_{\mbox{\tiny {\rm BP}}}(\Psi_{0})$ 
between the two electrons in the bipolaron grows 
more and more as $\eta$ turns out to be closer to $1$ 
and also that the coupling constant $\alpha$ 
in the weak-coupling regime 
cannot stem its growth. 
That is not the case in the strong-coupling 
regime (see Sec.\ref{sec:SLSCR}). 
This is a noticeable difference between 
the cases in the weak-coupling regime 
and in the strong-coupling one. 
 
Moreover, this growth tells us that if a bipolaron 
has a ground state $\Psi_{0}$ for $\eta$ sufficiently close to $1$, 
then $\overline{B}(d_{\mbox{\tiny {\rm BP}}}(\Psi_{0}))$ 
completely contains the whole crystal. 
On the other hand, since the bipolaron 
in the ground state $\Psi_{0}$ should be 
formed in the crystal, the two electrons in the bipolaron 
must exist in the crystal. Thus, we have met a contradiction. 
Therefore, reductio ad absurdum makes us conclude from these 
two facts contradicting each other that {\it any bipolaron in 
a ground state cannot be formed in the crystal 
if $\eta$ is close to $1$ 
in the weak-coupling regime.} 
We will give another lower bound to 
$d_{\mbox{\tiny {\rm BP}}}(\Psi_{0})$ 
in the inequality 
(\ref{eq:estimate-bipolaron-radius-0}) later. 

Our above result also states, in particular, 
that {\it the relative motion of the bipolaron 
in a ground state 
is not spatially localized in $\overline{B}(r)$, 
provided that the ground state 
exists under the condition:} 
\begin{eqnarray}
E_{\mathrm {w}}(\alpha) < \frac{U}{r}\qquad 
\left(\mbox{i.e.},\,\,\, 0 \le 1 - 
\frac{\Lambda}{8r} < \eta
\right).  
\label{eq:assumption}
\end{eqnarray}  
In terms of the binding energy 
with Eq.(\ref{eq:emin}), 
the condition (\ref{eq:assumption}) means  
\begin{eqnarray}
\nonumber 
-\, (2E_{\mbox{\tiny {\rm SP}}} 
- E_{\mbox{\tiny {\rm BP}}}) 
&\approx& -\, E_{\mathrm {w}}(\alpha) - 
E_{\mbox{\tiny {\rm BP}}} + \frac{U}{r} \\ 
&\ge& -\, E_{\mathrm {w}}(\alpha)  + \frac{U}{r} 
= \sqrt{2}\,\alpha
\left(\frac{1}{1-\eta} - \frac{8r}{\Lambda}\right)
\frac{1}{r} > 0
\label{negative-r-binding}
\end{eqnarray}
by estimating $E(r)$ at 
$E_{\mathrm {w}}(\alpha) + E_{\mbox{\tiny {\rm BP}}}$. 
Thus, as for the relation between the spatial localization 
of a ground state and the binding energy, 
we can interpret our result as that 
{\it a ground state of the bipolaron with 
the size of $d_{\mbox{\tiny {\rm BP}}}(\Psi_{0})$ 
less than $r$ satisfying the condition (\ref{eq:assumption}) 
cannot exist because of 
the negative binding energy (\ref{negative-r-binding})}. 

We define the number of phonons in a ground state 
$\Psi_{0}$ by $N_{\mathrm {ph}}(\Psi_{0}) 
:= \big<\Psi_{0}\big|\Psi_{0}\big>^{-1}
\big<\Psi_{0}\big|\sum_{k}a_{k}^{\dagger}a_{k}
\big|\Psi_{0}\big>$. 
Then, it is shown in this section 
that {\it  $N_{\mathrm {ph}}(\Psi_{0})$ 
is estimated as:} 
\begin{eqnarray}
\frac{U}{d_{\mbox{\tiny {\rm BP}}}(\Psi_{0})} 
+ (2E_{\mbox{\tiny {\rm SP}}} - 
E_{\mbox{\tiny {\rm BP}}})
\le N_{\mathrm {ph}}(\Psi_{0}) 
\le 
2E_{\mathrm {w}}(\alpha) + E_{\mbox{\tiny {\rm BP}}} 
+ 2\sqrt{
E_{\mathrm {w}}(\alpha)
(
E_{\mathrm {w}}(\alpha) + E_{\mbox{\tiny {\rm BP}}} 
)
}. 
\label{eq:estimate-phon-numbers}
\end{eqnarray}
The first inequality yields {\it the upper bound 
to the binding energy:} 
\begin{eqnarray}
2E_{\mbox{\tiny {\rm SP}}} - 
E_{\mbox{\tiny {\rm BP}}} 
\le N_{\mathrm {ph}}(\Psi_{0}) 
- \frac{U}{d_{\mbox{\tiny {\rm BP}}}(\Psi_{0})}  
< N_{\mathrm {ph}}(\Psi_{0}). 
\label{eq:up-bound-binding-energy}
\end{eqnarray}
Hence it follows that 
{\it another lower bound to 
$d_{\mbox{\tiny {\rm BP}}}(\Psi_{0})$ 
is given as:}
\begin{eqnarray}
\frac{U}{N_{\mathrm {ph}}(\Psi_{0}) 
- (2E_{\mbox{\tiny {\rm SP}}} - 
E_{\mbox{\tiny {\rm BP}}})} 
\le 
d_{\mbox{\tiny {\rm BP}}}(\Psi_{0}).  
\label{eq:estimate-bipolaron-radius-0}
\end{eqnarray}

Let us begin by proving the first result. 
We adopt Lieb's idea\cite{lieb} into our argument 
to prove it. 
Here we assume that $\widetilde{H}_{\mbox{\tiny BP}}$ 
has a bound state $\Psi_{n}$ with its eigenenergy $E$, i.e., 
$\widetilde{H}_{\mbox{\tiny BP}}\Psi_{n} 
= E\Psi_{n}$, and that it is spatially 
localized in $\overline{B}(r)$. 
Here we can suppose $\Psi_{n}$ is 
normalized, i.e.,  $\big<\Psi_{n}\big|
\Psi_{n}\big> = 1$, without any loss of generality. 
Because we only have to take 
$\big<\Psi_{n}\big|\Psi_{n}\big>^{-1/2}\Psi_{n}$ 
as the normalized bound state in case that 
$\Psi_{n}$ is not normalized. 
Then, since $P_{1}^{2}/2M_{1} \ge 0$, 
evaluating RHS of 
$0 = 
\big<|X_{2}|\Psi_{n}\big| 
\widetilde{H}_{\mbox{\tiny BP}} - E 
\big|\Psi_{n}\big>$ from below yields  
\begin{eqnarray*}
0 \ge 
\frac{1}{2M_{2}}
\big<|X_{2}|\Psi_{n}\big|
P_{2}^{2}\Psi_{n}\big> 
+ U 
+
\big<|X_{2}|^{1/2}\Psi_{n}\big|
H_{\mathrm {ph}} 
+ \widetilde{H}_{{\mathrm {el}}\mbox{-}{\mathrm {ph}}}
\big| |X_{2}|^{1/2}\Psi_{n}\big>   
- E
\big<\Psi_{n}\big| |X_{2}|
\big| \Psi_{n}\big>. 
\end{eqnarray*}
This leads us to the inequality:  
\begin{eqnarray}
0 \ge \frac{1}{2M_{2}} 
\big<|X_{2}|\Psi_{n}\big|P_{2}^{2}\Psi_{n}\big> 
+ U - \left(E+E_{\mathrm {w}}(\alpha)\right)r,  
\label{eq:(III-3)}
\end{eqnarray}
where we used the spatial localization of $\Psi_{0}$ 
and the fact that $E+E_{\mathrm {w}}(\alpha) 
\ge E_{\mbox{\tiny {\rm BP}}}
+E_{\mathrm {w}}(\alpha) \ge 0$ 
by the inequality (\ref{eq:lower-bound-weak}). 
It follows from the same argument that 
\begin{eqnarray}
0 \ge 
\frac{1}{2M_{2}}
\Big<P_{2}^{2}\Psi_{n}\Big|
|X_{2}|\Psi_{n}\Big> 
+ U  
- \left( E+E_{\mathrm {w}}(\alpha)\right)r. 
\label{eq:(III-3')}
\end{eqnarray}
Since $\Delta_{X_{2}}|X_{2}|^{-1} = 
4\pi\delta(X_{2})$, we have 
$\left(P_{2}^{2}|X_{2}| + 
|X_{2}|P_{2}^{2}\right)/2 
=  
|X_{2}|P_{2}|X_{2}|^{-1}P_{2}|X_{2}| 
- 2\pi |X_{2}|\delta(X_{2})|X_{2}|$. 
Thus, we reach the equation and inequality: 
\begin{eqnarray}
\frac{1}{2M_{2}}
\Big<\Psi_{n}\Big|\, P_{2}^{2}|X_{2}| 
+ |X_{2}|P_{2}^{2}\, \Big|\Psi_{n}\Big> 
=  
\frac{1}{M_{2}}
\Big< P_{2}|X_{2}|\Psi_{n}\Big|\, 
|X_{2}|^{-1}\, \Big| P_{2}|X_{2}|\Psi_{n}\Big> 
\ge 0. 
\label{eq:(III-4)}
\end{eqnarray}
Combining the inequalities 
(\ref{eq:(III-3)})--(\ref{eq:(III-4)}), 
we arrive at the first desired 
inequality (\ref{eq:assumption-n}).

To show the inequality 
(\ref{eq:lower-estimate-bipolaron-radius}), 
we use reductio ad absurdum. 
As an assumption for it 
we suppose $d_{\mbox{\tiny {\rm BP}}} 
< d_{\mbox{\tiny {\rm BP}}}^{\mathrm {low}}$.
Then, there is a positive number $r'$ so that 
$d_{\mbox{\tiny {\rm BP}}} 
< r' < d_{\mbox{\tiny {\rm BP}}}^{\mathrm {low}}$. 
Since $d_{\mbox{\tiny {\rm BP}}} < r'$, 
the above fact already proved 
says $E_{\mathrm {w}}(\alpha) + E_{\mbox{\tiny {\rm BP}}} 
\ge U/r'$, and so $r' \ge 
d_{\mbox{\tiny {\rm BP}}}^{\mathrm {low}}$. 
This is not consistent with the fact that 
$r' < d_{\mbox{\tiny {\rm BP}}}^{\mathrm {low}}$. 
Thus, reductio ad absurdum supplies the second 
desired inequality.

Let us proceed with the proof of our
next statement. 
Since $E_{\mbox{\tiny BP}} \le 0$, 
the simple term 
$E_{\mathrm {w}}(\alpha)$ can substitute 
for $E_{\mathrm {w}}(\alpha) 
+ E_{\mbox{\tiny BP}}$ 
in the inequality (\ref{eq:assumption-n}) 
for the ground state $\Psi_{0}$. 
Thus, the contraposition of the first result 
leads us to the third desired result 
on the spatial nonlocalization.

Let us estimate the number of phonons in 
a ground state $\Psi_{0}$ now. 
We regard $\Psi_{0}$ as being normalized. 
We define the Hamiltonian $\sum_{j}
\widetilde{H}^{(j)}_{\mbox{\tiny {\rm SP}}}$ 
of the two separate single polarons 
for $\widetilde{H}_{\mbox{\tiny {\rm BP}}}$
by using the coordinates $X_{1}$ and $X_{2}$, i.e., 
$\sum_{j}\widetilde{H}^{(j)}_{\mbox{\tiny {\rm SP}}} 
:= \sum_{j=1,2}P_{j}^{2}/2M_{2} + 2H_{\mathrm {ph}} 
+ \widetilde{H}_{{\mathrm {el}}\mbox{-}{\mathrm {ph}}}$.  
In addition, we note that the Hamiltonian of the 
two separate single polarons for 
$\widetilde{H}_{\mbox{\tiny {\rm BP}}}^{\mathrm {rel}}$ 
is $\sum_{j}^{\mathrm {rel}}\widetilde{H}^{(j)}_{
\mbox{\tiny {\rm SP}}} = 
P_{2}^{2}/2M_{2} + 2H_{\mathrm {ph}} 
+ \widetilde{H}_{{\mathrm {el}}\mbox{-}{\mathrm {ph}}}
(\rho_{Q})$, and that for 
$\widetilde{H}_{\mbox{\tiny {\rm BP}}}^{\Pi=0}$ 
is 
$\sum_{j}^{\Pi=0}\widetilde{H}^{(j)}_{
\mbox{\tiny {\rm SP}}} = 
P_{2}^{2}/2M_{2} + 2H_{\mathrm {ph}} 
+ P_{\mathrm {ph}}^{2}/2M_{1} 
+ \widetilde{H}_{{\mathrm {el}}\mbox{-}{\mathrm {ph}}}
(\rho_{O})$.

Note $1 = |X_{2}|^{-1/2}|X_{2}|^{1/2}$ now. 
Then, Schwarz's inequality implies that $1 = 
\big<|X_{2}|^{-1/2}\Psi_{0}\big||X_{2}|^{1/2}\Psi_{0}\big> 
\le \big<\Psi_{0}\big| 
|X_{2}|^{-1}\big|\Psi_{0}\big>^{1/2}
\big<\Psi_{0}\big| 
|X_{2}|\big|\Psi_{0}\big>^{1/2}$, 
making the inequality 
$1/d_{\mbox{\tiny {\rm BP}}}(\Psi_{0}) 
\le \big<\Psi_{0}\big| 
|X_{2}|^{-1}\big|\Psi_{0}\big>$. 
Hence it follows from this inequality, 
together with 
sandwiching the equation 
$\sum_{j}\widetilde{H}_{\mbox{\tiny {\rm SP}}}^{(j)} 
- 2E_{\mbox{\tiny {\rm SP}}} 
= \widetilde{H}_{\mbox{\tiny {\rm BP}}} 
- 2E_{\mbox{\tiny {\rm SP}}} 
+ H_{\mathrm {ph}} - U/|X_{2}|$ between 
$\big<\Psi_{0}\big|$ and $\big|\Psi_{0}\big>$, 
that 
$0 \le 
(E_{\mbox{\tiny {\rm BP}}} - 2E_{\mbox{\tiny {\rm SP}}}) 
+ N_{\mathrm {ph}}(\Psi_{0}) 
- U/d_{\mbox{\tiny {\rm BP}}}(\Psi_{0})$. 
Therefore,  the lower bound in the inequality 
(\ref{eq:estimate-phon-numbers}) is obtained 
from this inequality. 
As for the upper bound, we use  
the simple operator inequality: 
$$C\widetilde{H}_{\mbox{\tiny {\rm BP}}} 
- H_{\mathrm {ph}} 
\ge 
(C-1)\left\{ 
H_{\mathrm {ph}} 
+ \frac{C}{(C-1)}
\widetilde{H}_{{\mathrm {el}}\mbox{-}{\mathrm {ph}}} 
\right\}
\ge 
-\, \frac{C^{2}}{C-1}E_{\mathrm {w}}(\alpha), 
\qquad C > 1,$$ 
to come up with 
$H_{\mathrm {ph}} \le 
C\widetilde{H}_{\mbox{\tiny {\rm BP}}} 
+ C^{2}E_{\mathrm {w}}(\alpha)/(C-1)$. 
Putting this inequality between 
$\big<\Psi_{0}\big|$ and $\big|\Psi_{0}\big>$ 
leads us to the inequality 
$N_{\mathrm {ph}}(\Psi_{0}) 
\le \inf_{C>1}\left[ 
CE_{\mbox{\tiny {\rm BP}}} 
+ C^{2}E_{\mathrm {w}}(\alpha)/(C-1)
\right]$. 
RHS of this inequality attains 
the lower bound in 
the inequality (\ref{eq:estimate-phon-numbers}).

\section{Bipolaron Formation}
\label{sec:formation}

In this section we deal with the original 
$H_{\mbox{\tiny {\rm BP}}}$ (i.e., in the case 
$\rho(x) \equiv 1$) in the strong-coupling regime 
and derive two effective Hamiltonians from 
$H_{\mbox{\tiny {\rm BP}}}$ 
by modifying Bogolubovs' method.\cite{BB} 
Our modified method is similar to Adamowski's\cite{adamowski} 
and ours.\cite{HHS} 
More precisely, 
we seek a canonical transformation 
$U_{\theta}$  with the parameter $\theta \ge 0$ 
so that the transformed Hamiltonian has the form 
of $H_{\mbox{\tiny {\rm BP}}}(\theta) 
:= U_{\theta}^{*}H_{\mbox{\tiny BP}}U_{\theta} 
= H_{{\mathrm {eff}}}(\theta) 
+ H_{\mathrm {ph}} 
+ 
H_{{\mathrm {el}}\mbox{-}{\mathrm {ph}}}(\theta)  
+ \Sigma_{\theta}$, 
where $H_{{\mathrm {eff}}}(\theta)$ is 
an effective Hamiltonian in quantum mechanics, 
and $\Sigma_{\theta}$ a divergent energy 
as $\theta\to\infty$.  
The canonical transformation $U_{\theta}$ 
requires that the effective Hamiltonian 
$H_{{\mathrm {eff}}}(\theta)$ 
should gain an attractive potential $V(\theta)$ 
from the phonon field as 
$H_{{\mathrm {eff}}}(\theta) = 
H_{{\mathrm {el}}\mbox{-}{\mathrm {el}}} 
+ V(\theta)$. 
Thanks to this extra attractive potential $V(\theta)$, 
we expect that there is a critical point $\theta_{c}$ 
so that the Hamiltonian $H_{{\mathrm {eff}}}(\theta)$ 
itself or the Hamiltonian 
$H_{{\mathrm {eff}}}^{\mathrm {rel}}(\theta)$ 
for the relative motion of 
$H_{{\mathrm {eff}}}(\theta)$ 
has a ground state if 
$\theta > \theta_{c}$. 
On the other hand, it has no ground state 
if $\theta < \theta_{c}$. 
Thus, we can expect that the bipolaron 
is stable at/near the point $\theta_{\mathrm {c}}$. 
Actually, according to the recent result,\cite{MS} 
if we apply the above method to 
$\widetilde{H}_{\mbox{\tiny {\rm BP}}}^{\Pi=0}$, 
then we might be able to show that 
$\widetilde{H}_{\mbox{\tiny {\rm BP}}}^{\Pi=0}$ 
has a ground state 
for sufficiently large $\theta > 0$.

To find the canonical transformation $U_{\theta}$, 
we adopt the notion of Feynman's virtual phonon\cite{feynman} 
and the image of Peeters and Devreese's 
classical bipolaron,\cite{PD} 
being led to the classical images of 
the two kinds of states of bipolaron: 
{\it balanced state} and {\it unbalanced state}. 
The balanced state designates the image 
that the virtual phonon sits at the center 
of two electrons. 
On the other hand, the unbalanced state 
represents the image that the virtual phonon 
is not on the line segment between the two electrons.   
Then, two effective Hamiltonians are derived 
in quantum mechanics through these classical images.

It is shown in this section that 
{\it the approximation 
(\ref{eq:FE}) yields an effective Hamiltonian 
describing balanced state in quantum mechanics:}
\begin{eqnarray}
H_{\mathrm {eff}}(\theta) 
= H_{\mathrm {el}\mbox{-}{\mathrm {el}}} 
+ V(\theta)  
= \frac{1}{2}p_{1}^{2} + \frac{1}{2}p_{2}^{2}
+ \frac{U(\theta)}{|x_{1}-x_{2}|}
\label{eq:eff-Hbal} 
\end{eqnarray}
{\it with} 
\begin{eqnarray}
U(\theta) 
= U - \sqrt{2}\,\alpha\theta 
= \sqrt{2}\,\alpha
\left( \frac{1}{1-\eta} - \theta\right).   
\label{eq:eff-U(theta)}
\end{eqnarray} 
It is clear that $\theta_{\mathrm {c}} = 1/(1-\eta)$.
Let us fix $c_{*}$ arbitrarily so that 
$0 < c_{*} < 1$. 
We set $c_{\mbox{\tiny {\rm BP}}}$ as:  
\begin{eqnarray}
c_{\mbox{\tiny {\rm BP}}} := 
\frac{2}{5}
\left( c_{*} 
- \frac{1}{\theta(1-\eta)}
\right)^{2}. 
\label{eq:cBP}
\end{eqnarray}
Then, using the effective Hamiltonian $H_{\mathrm {eff}}(\theta)$ 
in Eq.(\ref{eq:eff-Hbal}), we show that 
{\it an upper bound to the ground state energy 
$E_{\mbox{\tiny {\rm BP}}}$ is given as:}
\begin{eqnarray}
E_{\mbox{\tiny {\rm BP}}} 
\le -\, c_{\mbox{\tiny {\rm BP}}}\alpha_{\theta}^{2} 
+ \frac{\sqrt{2}\,\alpha_{\theta}}{\pi}
\left( 4K + \frac{K^{3}}{3}\right),   
\label{eq:upper-bound-to-gse}
\end{eqnarray}
{\it provided that}
\begin{eqnarray}
1 - \frac{1}{c_{*}\theta} > \eta.
\label{eq:katei}
\end{eqnarray}
We note that the condition (\ref{eq:katei}) 
prohibits us from taking the limit $\theta\to 0$ 
in the inequality (\ref{eq:upper-bound-to-gse}) 
because 
$$\theta > \frac{1}{c_{*}(1-\eta)} > \frac{1}{1-\eta}.$$ 

Meanwhile, it is also shown in this section that 
{\it a lower bound to 
$E_{\mbox{\tiny {\rm BP}}}$ as:}
\begin{eqnarray} 
E_{\mbox{\tiny {\rm BP}}} 
\ge 
\left(
\inf_{\varphi}{\mathcal E}_{\theta}(\varphi)
\right)
\alpha_{\theta}^{2} 
- \big|1 - \frac{1}{\theta}\big|
\frac{4\sqrt{2}\,\alpha_{\theta}}{\pi}K,  
\label{eq:gse-functional}
\end{eqnarray}
{\it where} 
\begin{eqnarray} 
\nonumber
{\mathcal E}_{\theta}(\varphi) 
&:=& 
\frac{1}{2}\int\int 
d^{3}x_{1}d^{3}x_{2}
\Biggl[
\big|\nabla_{x_{1}}\varphi(x_{1},x_{2})\big|^{2} 
+ \big|\nabla_{x_{2}}\varphi(x_{1},x_{2})\big|^{2} 
+ \frac{2\sqrt{2}}{\theta (1-\eta)}
\frac{|\varphi(x_{1},x_{2})|^{2}}{|x_{1}-x_{2}|} 
\Biggr] \\ 
&{}& 
- \frac{1}{\sqrt{2}}
\int\int d^{3}x_{1}d^{3}x_{2}
\int\int d^{3}y_{1}d^{3}y_{2} 
\sum_{j,j'=1,2} 
\frac{\big|\varphi(x_{1},x_{2})\big|^{2}
\big|\varphi(y_{1},y_{2})\big|^{2}}{|x_{j}-y_{j'}|}
\label{eq:VG-functional}
\end{eqnarray}
{\it is an energy functional describing unbalanced state 
and $\inf_{\varphi}{\mathcal E}_{\theta}(\varphi) < 0$.} 
This functional is the almost same as 
Vinetskii and Giterman's.\cite{VG,AM,lakhno,SD} 
Namely, for wave functions $\varphi(x_{1},x_{2})$ 
satisfying $\varphi(x_{1},x_{2}) = \varphi(x_{2},x_{1})$, 
our energy functional becomes theirs.  
Here, we note that the estimate (\ref{eq:gse-functional}) 
in the strong-coupling regime recovers the estimate 
(\ref{eq:lower-bound-weak}) in the weak-coupling one 
as $\theta = 0$. 

At the end of this section we show that 
{\it the approximation (\ref{eq:FE}) yields another 
effective Hamiltonian describing unbalanced state 
in quantum mechanics:} 
\begin{eqnarray} 
H_{\mathrm {eff}}(\theta)  
= 
\alpha_{\theta}^{2}
\Biggl[
\sum_{j=1,2}\frac{1}{2}p_{j}^{2} 
- \sqrt{2}
\sum_{j=1,2}\frac{1}{|x_{j}|} 
+ \frac{\sqrt{2}}{\theta(1-\eta)|x_{1}-x_{2}|}
\Biggr].  
\label{eq:eff-Hunbal}
\end{eqnarray}
Here we note that the approximation (\ref{eq:FE}) 
breaks the translation invariance 
so that the effective Hamiltonian $H_{\mathrm {eff}}(\theta)$ 
in Eq.(\ref{eq:eff-Hunbal}) describes that the virtual phonon 
is nailed down at the origin. 
Let us set $E_{\mathrm {s}}(\alpha)$ as:  
\begin{eqnarray}
E_{\mathrm {s}}(\alpha) 
:= 
\left( \frac{2\sqrt{2}}{r} -\, \frac{1}{r^{2}} - 1
\right)\alpha\theta  
\label{eq:E(r)-existence}
\end{eqnarray} 
now. 
Then, {\it the effective Hamiltonian $H_{\mathrm {eff}}(\theta)$ 
in Eq.(\ref{eq:eff-Hunbal}) has a ground state provided 
that there is an $r > 0$ 
so that} 
\begin{eqnarray}
\frac{U}{r} < E_{\mathrm {s}}(\alpha).  
\label{eq:existence}
\end{eqnarray} 
Thus, we are led to the conclusion that 
$\theta_{\mathrm {c}} \le 1/(2-\sqrt{2})(1-\eta)$. 
The condition (\ref{eq:existence}) puts restrictions 
on $\theta, \eta$ and $r$. 
Namely, {\it the sufficient condition for 
the inequality (\ref{eq:existence}) is:}
\begin{eqnarray}
1- \frac{1+\sqrt{2}}{\sqrt{2}\,\theta} > \eta,  
\label{eq:restrictions}
\end{eqnarray}
{\it and} 
\begin{eqnarray}
R_{\theta,\eta} - \sqrt{{R_{\theta,\eta}}^{2} - 1} 
< r 
< R_{\theta,\eta} + \sqrt{{R_{\theta,\eta}}^{2} - 1},  
\label{eq:r-estimates}
\end{eqnarray}
{\it where} 
$$R_{\theta,\eta} 
:= \sqrt{2}
\left(
1 - \frac{1}{2\theta(1-\eta)}
\right).$$ 
Therefore, for sufficiently large $\theta$ 
(i.e., $\theta \approx \infty$), 
$0.585 \le r/r_{{}_{\mathrm {fp}}} 
\le 3.415$.

Since the inequalities (\ref{theorem:BE1-2}), 
(\ref{eq:upper-bound-to-gse}) 
and (\ref{eq:gse-functional}) say 
that $E(r) \ge 0$ is dominated 
from below by $\sqrt{2}\,\alpha_{\theta}/r$ 
and/or $E_{\mathrm {s}}(\alpha)$ for sufficiently 
large $\theta > 0$, 
in the case of the balanced state 
the condition $\theta > 1/(1-\eta)$ 
makes the estimated binding energy positive: 
\begin{eqnarray}
-\, (2E_{\mbox{\tiny {\rm SP}}} 
- E_{\mbox{\tiny {\rm BP}}}) 
\lessapprox 
- \frac{\sqrt{2}\,\alpha_{\theta}}{r} + \frac{U}{r} 
= 
\sqrt{2}\,\alpha
\left(
\frac{1}{1-\eta} - \theta
\right)\frac{1}{r} < 0. 
\label{eq:appearance-theta1}
\end{eqnarray}
In the case of the unbalanced state 
the condition (\ref{eq:existence}) 
also makes the estimated binding energy positive:  
\begin{eqnarray}
-\, (2E_{\mbox{\tiny {\rm SP}}} 
- E_{\mbox{\tiny {\rm BP}}}) 
\lessapprox 
- E_{\mathrm {s}}(\alpha) + \frac{U}{r} 
= 
\alpha_{\theta}
\left[
1 + \frac{1}{r^{2}} 
+ \sqrt{2}\left(
\frac{1}{\theta(1-\eta)} - 2
\right)\frac{1}{r}
\right] < 0. 
\label{eq:appearance-theta2}
\end{eqnarray}
Thus, the appearance of $\theta$ in 
Eqs.(\ref{eq:appearance-theta1}) and 
(\ref{eq:appearance-theta2}) is very different 
than it does not appear in Eqs.(\ref{eq:Intr-BE}) and 
(\ref{negative-r-binding}).

\subsection{Strategy for Effective Hamiltonians} 
\label{subsec:strategy}

In Ref.\onlinecite{HHS} we derived 
the Coulomb attractive potential 
from the electron-boson interaction 
for the model introduced by Gross.\cite{gross} 
Similarly, we use such a canonical transformation 
with trial functions $\beta_{j}(k,x_{j})$, 
$j = 1, 2$, which have the parameter $\theta$.  
By controlling the trial functions, 
we derive the effective 
Hamiltonian $H_{{\mathrm {eff}}}(\theta)$. 
Set $\beta_{j,k}$ as $\beta_{j,k} 
:= \beta_{j}(k,x_{j})$, where 
we assume $\beta_{j}(k,x_{j})^{*} 
= \beta_{j}(-k,x_{j})$ for every $x_{j}$ and 
$j = 1, 2$, i.e., 
$\beta_{j,k}^{*} = \beta_{j,-k}$. 
We give the generator $T$ by 
$T := i \sum_{j=1,2}\sum_{k}
\{ V_{k}\beta_{j,k}e^{ik\cdot x_{j}}a_{k} 
- 
V_{k}^{*}\beta_{j,k}^{*}e^{-ik\cdot x_{j}}a^{\dagger}_{k}
\}$. 
Then, we obtain the canonical 
transformation $U_{\theta}$ 
by $U_{\theta} := e^{iT}$. 
Precisely writing down, we obtain each term 
in the following. 
The effective potential 
$V(\theta)$ is given by 
$$
V(\theta) := 
2\sum_{k}|V_{k}|^{2}
\Bigg\{ 
\beta_{1,k}^{*}\beta_{2,k} 
- \Big( \beta_{1,k}^{*} + \beta_{2,k}\Big)
\Bigg\}e^{-ik(x_{1}-x_{2})}.
$$
The electron-phonon interaction 
$H_{{\mathrm {el}}\mbox{-}{\mathrm {ph}}}(\theta)$ 
is decomposed into the two parts as 
$H_{{\mathrm {el}}\mbox{-}{\mathrm {ph}}}(\theta) 
= H_{{\mathrm {el}}\mbox{-}{\mathrm {ph}}}^{(1)}  
+ H_{{\mathrm {el}}\mbox{-}{\mathrm {ph}}}^{(2)}$, 
which are respectively defined by   
\begin{eqnarray*} 
&{}& 
H_{{\mathrm {el}}\mbox{-}{\mathrm {ph}}}^{(1)} 
:= 
\sum_{j=1,2}\sum_{k}
\Bigg\{
V_{k}\left( 
1 - \beta_{j,k} 
- \frac{1}{2}D\left(\beta_{j,k}\right)
\right) e^{ikx_{j}}a_{k} \\ 
&{}& \qquad\qquad\qquad\qquad
+ 
V_{k}^{*}\left( 
1 - \beta_{j,k}^{*} 
- \frac{1}{2}D\left(\beta_{j,k}\right)^{*}
\right) e^{-ikx_{j}}a^{\dagger}_{k} 
\Bigg\}, \\ 
&{}& 
H_{{\mathrm {el}}\mbox{-}{\mathrm {ph}}}^{(2)} 
:= 
\sum_{j=1,2}
\left\{ p_{j}A_{j} + A_{j}^{\dagger}p_{j}\right\} 
+ 
\frac{1}{2}
\sum_{j=1,2}
\left\{ A_{j}^{2} + 2A_{j}^{\dagger}A_{j} 
+ A_{j}^{\dagger\, 2}\right\}, 
\end{eqnarray*}
where 
$D(\beta_{j,k}) = - \Delta_{x_{j}}\beta_{j,k} 
- 2ik\nabla_{x_{j}}\beta_{j,k} 
+ k^{2}\beta_{j,k}$ 
and $A_{j} = \sum_{k}V_{k}
\Big( 
k\beta_{j,k} - i (\nabla_{x_{j}}\beta_{j,k})
\Big) 
e^{ikx_{j}}a_{k}$. 
The divergent energy as $\theta\to\infty$ 
is 
\begin{eqnarray*} 
\Sigma_{\theta} 
&=& 
\sum_{k}|V_{k}|^{2}
\Bigg\{
\left( 1 + \frac{k^{2}}{2}\right)
\left( 
|\beta_{1,k}|^{2} + |\beta_{2,k}|^{2}
\right) 
+ 
\frac{1}{2}
\left(
|\nabla_{x_{1}}\beta_{1,k}|^{2}
+ |\nabla_{x_{2}}\beta_{,k}|^{2}
\right)  \\ 
&{}& \qquad\qquad\qquad 
- \Re\left(\beta_{1,k} + \beta_{2,k}\right)
\Bigg\}.
\end{eqnarray*}

The main problem in our strategy is what is 
the principle to find the best effective 
Hamiltonian. 
Here we check well-known method for a while 
by way of trial.  
If we employ $\beta_{j}(k) = \theta(1 + k^{2}/2)^{-1}$ 
for $\theta > 0$ ($j = 1, 2$), 
the $e^{iT}$ as $\theta = 1$ 
is the very unitary operator we used\cite{HHS} 
for the model introduced by Gross. 
That is, it is the unitary operator 
in Tomonaga's intermediate coupling approximation\cite{tomonaga} 
rearranged in Lee, Low, and Pines' study\cite{LLP,LP1} 
of the polaron. 
In the intermediate coupling approximation, 
the variational principle works so that 
the self-energy  turns out lowest. 
Thus, employing this $\beta_{j,k} 
\equiv \beta_{j}(k)$, 
with the help of 
the approximation of the Fourier expansion 
(\ref{eq:FE}) we obtain
\begin{eqnarray*}
V(\theta) 
= -\, \frac{4\sqrt{2}\pi\alpha}{\sqrt{(2\pi)^{3}}}
\left\{
\sqrt{\frac{\pi}{2}}
\frac{\theta(2 - \theta)}{|x_{1}-x_{2}|} 
-\, \theta(2 - \theta)\sqrt{\frac{\pi}{2}}\,\,
\frac{e^{-\sqrt{2}{|x_{1}-x_{2}|}}}{|x_{1}-x_{2}|} 
+  \theta^{2}\sqrt{\frac{\pi}{2}}\,\,
e^{-\sqrt{2}{|x_{1}-x_{2}|}}
\right\}. 
\end{eqnarray*}
Since $U =\sqrt{2}\,\alpha/(1-\eta)$, 
the effect Hamiltonian $H_{\mathrm {eff}}(\theta)$ 
can be derived as:  
\begin{eqnarray*}
H_{\mathrm {eff}}(\theta) 
&=&  
\frac{1}{2}p_{1}^{2} + \frac{1}{2}p_{2}^{2} 
+ \left( \frac{1}{1-\eta} - 
\theta(2 - \theta)\right)
\frac{\sqrt{2}\,\alpha}{|x_{1}-x_{2}|} \\ 
&{}&\qquad\qquad 
+ \theta(2 - \theta)\sqrt{2}\,\alpha 
\frac{e^{-\sqrt{2}|x_{1}-x_{2}|}}{|x_{1}-x_{2}|} 
-\, \theta^{2}\sqrt{2}\,\alpha 
e^{-\sqrt{2}|x_{1}-x_{2}|}.
\end{eqnarray*}
Unfortunately, 
we cannot gain enough attraction from 
the electron-phonon interaction because of 
the same reason as in Eq.(\ref{eq:Intr-BE}). 
Thus, we have now lost such a principle to derive 
the best effective Hamiltonian as in Lee, Low, 
and Pine's study. 
Then, we depend on the picture given by 
the combination of the two notions: 
Feynman's of virtual phonon,\cite{feynman} 
and Peeters and Devreese's of the classical 
bipolaron.\cite{PD} 
As is in Kadanoff's explanation,\cite{kadanoff} 
Feynman considered that the electron 
in the phonon cloud, which forms a single polaron, 
is coupled to another particle with a harmonic oscillator. 
We call the particle a (classical) virtual phonon. 
We use the notion of this virtual phonon. 
Although Feynman assumed harmonic oscillator 
for the interaction between the electron and 
the virtual phonon, 
we assume the Coulomb attractive 
potential between each electron and the virtual phonon 
instead as follows: 
We imagine the Peeters and Devreese's classical 
bipolaron first. 
We assume a (classical) virtual phonon sits at 
the center $(x_{1}+x_{2})/2$ of the line 
segment of $x_{1}$ and $x_{2}$. 
In this case, the two electrons can feel an attraction 
between themselves besides the Coulomb repulsion 
without noticing the existence of the virtual phonon, 
though the attraction is actually made by the virtual phonon. 
Thus, we reach the image of the balanced state. 
On the other hand, for the unbalanced state, 
the two electrons have to become aware of the existence 
of the virtual phonon.

\subsection{Effective Hamiltonian for Balanced State}
\label{subsec:balanced-state}

In this subsection, we consider how 
we can obtain the Coulomb attraction 
between the two electrons in quantum theory 
for the balanced state. 
To derive such the Coulomb attraction, 
we employ $\beta_{j,k}$ simply defined by 
$\beta_{1,k} = - \beta_{2,k} = 
\sqrt{\theta}$. 
Then, $\Sigma_{\theta}$ is approximated to 
$2\sqrt{2}\alpha(\theta-\sqrt{\theta})K/\pi 
+ \sqrt{2}\,\alpha\theta K^{3}/6\pi$ 
by the estimate (\ref{eq:(III-0)}), 
so that $\lim_{\theta\to\infty}\Sigma_{\theta}/\theta^{2} 
= 0$. 
Then, the approximation of the Fourier expansion 
(\ref{eq:FE}) leads us to the more effective 
$V(\theta)$:
\begin{eqnarray*}
V(\theta) 
= 
 -\, \frac{\sqrt{2}\,\alpha\theta}{|x_{1}-x_{2}|}.  
\end{eqnarray*}
Therefore, we obtain the effective Hamiltonian 
$H_{\mathrm {eff}}(\theta)$ defined in Eq.(\ref{eq:eff-Hbal}) 
representing the balanced state. 
Using coordinates of the center-of-mass and 
the relative motion, we realize that 
$H_{\mathrm {eff}}(\theta)$ 
is unitary-equivalent to 
$\widetilde{H}_{\mathrm {eff}}(\theta)$: 
$$\widetilde{H}_{\mathrm {eff}}(\theta) 
:= \frac{1}{4}P_{1}^{2} + 
\widetilde{H}^{\mathrm {rel}}_{\mathrm {eff}}(\theta),$$ 
where 
$$\widetilde{H}^{\mathrm {rel}}_{\mathrm {eff}}(\theta) 
:= P_{2}^{2} + \frac{U(\theta)}{|X_{2}|}.$$

At the end of this subsection, we show 
the upper bound (\ref{eq:upper-bound-to-gse}) 
to $E_{\mbox{\tiny {\rm BP}}}$. 
We now introduce a mathematical 
parameter $R > 2$ into $H_{\mbox{\tiny BP}}$ 
so that $\lim_{R\to\infty}H_{\mbox{\tiny BP}}(R) 
= H_{\mbox{\tiny BP}}$ (in the norm resolvent 
sense\cite{RS}). 
To introduce the parameter $R > 2$, 
we use a real-valued function 
$\gamma_{2}(x)$ defined by 
by $\gamma_{2}(x) := \chi_{R}(|x|)$, 
where $\chi_{R}(r) :=1$ for $r < R/2$ and 
$\chi_{R}(r) := 0$ for $r > R$ with linear interpolation. 
Then, it is easy to show that $\sup_{x}|\nabla_{x}
\gamma_{2}(x)| \le 2/R < 1$.  
We define the Hamiltonian 
$H_{\mbox{\tiny BP}}(R)$ with the parameter 
$R > 2$ by 
$H_{\mbox{\tiny {\rm BP}}}(R) :=
H_{{\mathrm {el}}\mbox{-}{\mathrm {el}}} + 
H_{\mathrm {ph}} 
+ H_{{\mathrm {el}}\mbox{-}{\mathrm {ph}}}(R)$, 
where $H_{{\mathrm {el}}\mbox{-}{\mathrm {ph}}}(R)$ 
is defined by putting $\gamma_{2}(x_{1}-x_{2})$ 
in front of $\sum_{j=1,2}\sum_{k}$ 
in the original $H_{{\mathrm {el}}\mbox{-}{\mathrm {ph}}}$ 
of $H_{\mbox{\tiny {\rm BP}}}$.

We can define the unitary-transformed 
Hamiltonian $H_{\mbox{\tiny {\rm BP}}}(\theta, R)$ 
by $H_{\mbox{\tiny {\rm BP}}}(\theta, R) 
:= e^{iT}H_{\mbox{\tiny BP}}(R)e^{-iT}$. 
Then, $H_{\mbox{\tiny {\rm BP}}}(\theta, R)$ 
is approximated as 
$H_{\mbox{\tiny {\rm BP}}}(\theta, R) 
\approx H_{{\mathrm {eff}}}(\theta, R) 
+ H_{\mathrm {ph}} 
+ 
H_{{\mathrm {el}}\mbox{-}{\mathrm {ph}}}(\theta,R)  
+ \Sigma_{\theta}(R)$ through the approximation (\ref{eq:FE}). 
Here the effective Hamiltonian 
$H_{\mathrm {eff}}(\theta,R)$ is given by 
$H_{\mathrm {eff}}(\theta,R) 
= H_{\mathrm {eff}}(\theta) 
+ {\mathrm {error}}_{{}_{R}}(x_{1}-x_{2},\theta)$ 
for the effective Hamiltonian $H_{\mathrm {eff}}(\theta)$ 
in Eq.(\ref{eq:eff-Hbal}) and the error term 
${\mathrm {error}}_{{}_{R}}(x_{1}-x_{2},\theta)$:   
$${\mathrm {error}}_{{}_{R}}(x_{1}-x_{2},\theta) 
= \Big( 
1 - \gamma_{2}(x_{1}-x_{2})^{2}
\Big)
\frac{\sqrt{2}\,\alpha\theta}{|x_{1}-x_{2}|},
$$ 
and the electron-phonon interaction 
$H_{{\mathrm {el}}\mbox{-}{\mathrm {ph}}}(\theta,R)$ is given by 
$H_{{\mathrm {el}}\mbox{-}{\mathrm {ph}}}(\theta,R)  
= H_{{\mathrm {el}}\mbox{-}{\mathrm {ph}}}^{(1)}(R)  
+ H_{{\mathrm {el}}\mbox{-}{\mathrm {ph}}}^{(2)}$ 
for $H_{{\mathrm {el}}\mbox{-}{\mathrm {ph}}}^{(2)}$ 
given in Sec.\ref{subsec:strategy} and 
$H_{{\mathrm {el}}\mbox{-}{\mathrm {ph}}}^{(1)}(R)$:  
\begin{eqnarray*}
H_{{\mathrm {el}}\mbox{-}{\mathrm {ph}}}^{(1)}(R) 
&:=&  
\sum_{j=1,2}\sum_{k}
\Bigg\{
V_{k}\left( 
\gamma_{2}(x_{1}-x_{2}) - \beta_{j,k} 
- \frac{1}{2}D\left(\beta_{j,k}\right)
\right) e^{ikx_{j}}a_{k}  \\ 
&{}& \qquad\qquad\qquad\qquad
+ 
V_{k}^{*}\left( 
\gamma_{2}(x_{1}-x_{2}) - \beta_{j,k}^{*} 
- \frac{1}{2}D\left(\beta_{j,k}\right)^{*}
\right) e^{-ikx_{j}}a^{\dagger}_{k} 
\Bigg\}. 
\end{eqnarray*}
The self-energy $\Sigma_{\theta}(R)$ 
in $H_{\mbox{\tiny {\rm BP}}}(\theta,R)$ is 
given by  
\begin{eqnarray*} 
&{}& 
\Sigma_{\theta}(R) = \sum_{k}|V_{k}|^{2}
\Bigg\{
\left( 1 + \frac{k^{2}}{2}\right)
\left( 
|\beta_{1,k}|^{2} + |\beta_{2,k}|^{2}
\right) 
+ 
\frac{1}{2}
\left(
|\nabla_{x_{1}}\beta_{1,k}|^{2}
+ |\nabla_{x_{2}}\beta_{,k}|^{2}
\right)  \\ 
&{}& \qquad\qquad\qquad 
- \gamma_{2}(x_{1}-x_{2})
\Re\left(\beta_{1,k} + \beta_{2,k}\right)
\Bigg\}.
\end{eqnarray*}
Using coordinates of the center-of-mass and 
the relative motion, we realize that 
$H_{\mathrm {eff}}(\theta,R)$ 
is unitary-equivalent to 
$\widetilde{H}_{\mathrm {eff}}(\theta,R) 
:= P_{1}^{2}/4 + 
\widetilde{H}^{\mathrm {rel}}_{\mathrm {eff}}(\theta,R)$, 
where 
$\widetilde{H}^{\mathrm {rel}}_{\mathrm {eff}}(\theta,R) 
:= \widetilde{H}^{\mathrm {rel}}_{\mathrm {eff}}(\theta) 
+ {\mathrm {error}}_{{}_{R}}(X_{2},\theta)$.  

For the real number $c_{*}$, we can take a unique point 
$r_{\mbox{\tiny {\rm BP}}}$ so that 
$C_{0}(r_{\mbox{\tiny {\rm BP}}}) 
= 1- c_{*}$, where $C_{0}(r)$ was defined 
in Eq.(\ref{eq:C_{0}}). 
We set $C_{2}(\mu, R)$ and $R_{\mbox{\tiny {\rm BP}}}$ 
as $C_{2}(\mu, R) := C_{0}(\mu R)$ 
and $R_{\mbox{\tiny {\rm BP}}} := 
r_{\mbox{\tiny {\rm BP}}}/\mu$, respectively. 
Then $C_{2}(\mu , R) \le 1 - c_{*}$, 
provided $R_{\mu} \le R$. 
Setting $r$ as $r = 1/\mu$ 
in the definition of $\psi_{\mu}(x_{j})$, 
we have the wave function $\phi_{\mu}(X_{1})$ 
as $\phi_{\mu}(X_{1}) := 
\sqrt{\mu^{3}/\pi}\,\, e^{-\mu |X_{1}|}$. 
We denote that of $X_{2}$ 
by $\psi_{\mu}(X_{2}) := \sqrt{\mu^{3}/\pi}
\,\, e^{-\mu |X_{2}|}$.  
Set $R_{\mbox{\tiny {\rm BP}}}$ as 
$R_{\mbox{\tiny {\rm BP}}} 
:= r_{\mbox{\tiny {\rm BP}}}/\mu$. 
Then $C_{2}(\mu , R) \le 1 - c_{*}$, 
provided $R_{\mbox{\tiny {\rm BP}}} \le R$.
It follows from simple estimates that 
\begin{eqnarray}
\big<\psi_{\mu}\big|\, 
\widetilde{H}_{\mathrm {eff}}^{\mathrm 
{rel}}(\theta,R)\, 
\big|\psi_{\mu}\big>  
\le 
\mu^{2} - \sqrt{2}\,\alpha\theta 
\left( c_{*} 
- \frac{1}{\theta(1-\eta)} \right)\mu
\label{eq:BP-BE1}  
\end{eqnarray}
for every $R$ which satisfies 
$R_{\mbox{\tiny {\rm BP}}} \le R$. 
It is easily shown that $\big<\phi_{\mu}\big|\, 
P_{1}^{2}\, \big|\phi_{\mu}\big> = \mu^{2}$. 
We define a wave function $\varphi(X_{1}, X_{2})$ 
by $\varphi(X_{1}, X_{2}) := \phi_{\mu}(X_{1})
\psi_{\mu}(X_{2})$.  
Then,  
for every $\mu > 0$ and $R \ge R_{\mbox{\tiny {\rm BP}}}$, 
the ground state energy $E_{\mbox{\tiny {\rm BP}}}(R) 
= \inf {\mathrm {Spec}}(H_{\mbox{\tiny {\rm BP}}}(\theta,R))$ 
of $H_{\mbox{\tiny {\rm BP}}}(\theta, R)$ 
is estimated from above as:  
\begin{eqnarray*} 
E_{\mbox{\tiny {\rm BP}}}(R) 
&\le& 
\big<\varphi\Omega_{\mathrm {ph}} \big|\, 
\widetilde{H}_{\mbox{\tiny {\rm BP}}}(\theta, R)\, 
\big|\varphi\Omega_{\mathrm {ph}}\big>  
\le 
\frac{1}{4}\big<\phi_{\mu}\big|\, 
P_{1}^{2}\, \big|\phi_{\mu}\big>
+ 
\big<\psi_{\mu}\big|\, 
\widetilde{H}_{\mathrm {eff}}^{\mathrm 
{rel}}(\theta,R)\, 
\big|\psi_{\mu}\big>  
+ \Sigma_{\theta}(R) \\ 
&\le& 
\frac{5}{4}
\left( \mu - \frac{2\sqrt{2}}{5}
\alpha\theta\left( 
c_{*} 
- \frac{1}{\theta(1-\eta)}
\right)\right)^{2} 
- \frac{2}{5}
\left( c_{*} 
- \frac{1}{\theta(1-\eta)}
\right)^{2}\alpha^{2}\theta^{2} \\ 
&{}& \qquad 
+ \frac{\sqrt{2}\,\alpha\theta}{\pi}
\left( 4K + \frac{K^{3}}{3}\right). 
\end{eqnarray*}
Setting the parameter $\mu$ 
as $\mu = \mu_{\mbox{\tiny {\rm BP}}} 
:= (2\sqrt{2}/5)
\alpha\theta (c_{*} 
- 1/\theta(1-\eta)) > 0$ 
because of the guarantee (\ref{eq:katei}) for it, 
we arrive at 
\begin{eqnarray*}
E_{\mbox{\tiny {\rm BP}}}(R) 
\le    
- c_{\mbox{\tiny {\rm BP}}}
\alpha_{\theta}^{2}
+ \frac{\sqrt{2}\,\alpha_{\theta}}{\pi}
\left( 4K + \frac{K^{3}}{3}\right)   
\end{eqnarray*} 
for every $R \ge {\mathrm {max}}
\{ 2, R_{\mbox{\tiny {\rm BP}}}\}$. 
Taking $R\to \infty$ in the above, 
we obtain our desired upper bound.

\subsection{Effective Hamiltonian for Unbalanced State}
\label{subsec:EHUS}

In this subsection, we consider how 
we can obtain the Coulomb attraction 
between each electron and virtual phonon 
in quantum theory for the unbalanced state.

Let us proceed with deriving the energy functional 
${\mathcal E}_{\theta}(\varphi)$ in 
Eq.(\ref{eq:VG-functional}) first. 
We arbitrarily fix a normalized phonon state $\Psi$ 
and a normalized 
wave function $\varphi(x_{1},x_{2})$ 
of the electron satisfying 
$\lim_{|x_{j}|\to\infty}\varphi(x_{1},x_{2}) = 0$. 
We denote by 
$\langle\quad\rangle$ 
an averaging over 
the wave function $\varphi(x_{1},x_{2})$. 
Set $U_{\alpha\theta} := 
U_{1,\alpha\theta}U_{2,\alpha\theta}$. 
Then, in the same way as in the case of single 
polaron, we can estimate the term 
$\langle U_{\alpha\theta}^{*}
H_{\mbox{\tiny BP}}
U_{\alpha\theta}\rangle$ in the equation,  
$$ 
\bigl<\left(U_{\alpha\theta}\varphi
\right)\Psi\big|\, 
H_{\mbox{\tiny BP}} 
\, 
\big|\left( U_{\alpha\theta}\varphi
\right)\Psi\bigr> 
= 
\Bigl<\Psi\Big| 
\langle U_{\alpha\theta}^{*}
H_{\mbox{\tiny BP}}
U_{j,\alpha\theta}
\rangle\, \Big|\Psi\Bigr>, 
$$
from below as:   
\begin{eqnarray*}
\langle U_{\alpha\theta}^{*}
H_{\mbox{\tiny BP}}
U_{\alpha\theta}\rangle  
&\ge& 
\frac{\alpha^{2}\theta^{2}}{2}
\left(
\langle p_{1}^{2}\rangle 
+ \langle p_{2}^{2}\rangle 
\right)
+ 
\langle 
\frac{\alpha\theta U}{|x_{1}-x_{2}|}
\rangle  
-\, \theta\sum_{k}|V_{k}|^{2}
\Bigl|\Bigl< 
e^{ik\cdot x_{1}/\alpha\theta} 
+ e^{ik\cdot x_{2}/\alpha\theta}
\Bigr>\Bigr|^{2} \\ 
&{}&  
-\, |\theta - 1|E_{\mathrm {w}}(\alpha).
\end{eqnarray*}
Using the approximation (\ref{eq:FE}), 
in the same way as for $E_{\mbox{\tiny SP}}$, 
we reach the lower bound 
(\ref{eq:gse-functional}).

To obtain an effective Hamiltonian 
for the unbalanced state, 
we only have to employ $G(\theta)$ as $T$. 
Namely, set $\beta_{j,k}$ as 
$\beta_{j,k} \equiv \beta_{j}(k) 
= e^{-ik\cdot x_{j}}\theta$. 
Then, we have $T = G(\theta)$. 
Under the approximation (\ref{eq:FE}), 
we approximate $H_{\mbox{\tiny {\rm BP}}}[\theta] 
:= U_{\alpha\theta}^{*}
e^{iG(\theta)}
H_{\mbox{\tiny {\rm BP}}}
e^{-iG(\theta)}
U_{\alpha\theta}$ 
to: 
\begin{eqnarray} 
\nonumber 
H_{\mbox{\tiny {\rm BP}}}[\theta]  
&\approx& 
H_{\mathrm {eff}}(\theta) + H_{\mathrm {ph}} \\ 
&{}& 
+ \sum_{j=1,2}\sum_{k}
\left\{
V_{k}\left( 
e^{ik\cdot x_{j}/\alpha\theta} 
- \theta\right) a_{k} 
+ 
V_{k}^{*}\left( 
e^{-ik\cdot x_{j}/\alpha\theta} 
- \theta\right) a^{\dagger}_{k} 
\right\} 
+ \Sigma_{\theta},   
\label{eq:app-ub-ham}
\end{eqnarray}
where $H_{\mathrm {eff}}(\theta)$ is given 
by Eq.(\ref{eq:eff-Hunbal}) and 
$\Sigma_{\theta} := \theta E_{\mathrm {w}}(\alpha)/4$. 
We note again that the approximation (\ref{eq:FE}) 
breaks the translation invariance in the original 
Hamiltonian $H_{\mbox{\tiny {\rm BP}}}$. 

Let us fix $r > 0$ arbitrarily. 
We denote by 
$\psi_{j}(x_{j})$ the wave functions 
$\sqrt{(1/\pi r^{3})}
\,\, e^{- |x_{j}|/r}$ for $j = 1,2$. 
We define the wave functions $\psi_{r}(x_{1}, x_{2})$ 
with the parameter $r$ by 
$\psi_{r}(x_{1}, x_{2}) := \psi_{1}(x_{1})\psi_{2}(x_{2})$. 
We decompose $H_{\mathrm {eff}}(\theta)$ into 
the following: 
\begin{eqnarray*}
H_{\mathrm {eff}}(\theta) 
&=& \alpha_{\theta}^{2}
\Biggl[
\frac{1}{2}p_{1}^{2} - \frac{\sqrt{2}}{|x_{1}|} 
+ \frac{1}{2}p_{1}^{2} 
- \left( 1 - \frac{1}{\theta(1-\eta)}\right)\frac{1}{|x_{1}|} \\ 
&{}& 
\qquad\qquad 
+ \frac{\sqrt{2}}{\theta(1-\eta)}
\left(\frac{1}{|x_{1}-x_{2}|} 
- \frac{1}{|x_{2}|}\right)
\Biggr].
\end{eqnarray*} 
It is well known (see page 89 of Ref.\onlinecite{RS4}) 
that $\big<\psi_{r} |\, |x_{1}-x_{2}|^{-1} - |x_{2}|^{-1}
\, |\psi_{r}\big> \le 0$ 
is obtained since $\psi_{1}(x_{1})$ is 
spherically symmetric. 
Thus, the inequality 
$\big<\psi_{r} |\, \alpha_{\theta}^{-2}
H_{\mathrm {eff}}(\theta)\, 
|\psi_{r}\big> \le 
\sqrt{2}(1/\theta(1-\eta) - 2)/r + 1/r^{2} < -1$ 
is obtained from the above decomposition and the inequality. 
Therefore, it follows from the HVZ theorem\cite{RS4} 
that $H_{\mathrm {eff}}(\theta)$ 
has a ground state under the condition (\ref{eq:existence}). 
The restrictions on $\theta$, $\eta$, and $r$ are obtained 
by solving the inequality (\ref{eq:existence}) with 
respect to the variable $r$.

\section{Spatial Localization in Strong-Coupling Regime}
\label{sec:SLSCR}

In this section we consider the approximated 
$H_{\mbox{\tiny {\rm BP}}}[\theta]$ 
given in Eq.(\ref{eq:app-ub-ham}). 
For a ground state $\Psi_{0}$ of 
$H_{\mbox{\tiny {\rm BP}}}[\theta]$, 
we define the radius $u_{\mbox{\tiny {\rm BP}}}
(\Psi_{0})$ of the sphere in which 
the two electron lives by 
\begin{eqnarray}
u_{\mbox{\tiny {\rm BP}}}(\Psi_{0}) 
:= 
\mathop{\mathrm {max}}_{j=1,2}
\left\{
\big<\Psi_{0}\big|\Psi_{0}\big>^{-1}
\big<\Psi_{0}\big|
|x_{j}|
\big|\Psi_{0}\big> 
\right\}.
\label{eq:u-radius-BP}
\end{eqnarray}
Then, we show in this section that 
{\it if the bipolaron has 
a ground state $\Psi_{0}$, then there is a relation:}
\begin{eqnarray}
u_{\mbox{\tiny {\rm BP}}}(\Psi_{0})
\ge  
\frac{1}{\sqrt{2}}
\left\{ 1 
+ \left( 1 + \frac{3\theta}{4}\right)
\frac{E_{\mathrm {w}}(\alpha)}{\alpha_{\theta}^{2}} 
\right\}^{-1}
\left( 
\frac{1}{\theta(1-\eta)}
- 2\right). 
\label{eq:SLSCR-result1}
\end{eqnarray}
Thus, we are led to the conclusion that, 
even if $\eta$ approaches $1$, 
$\theta > 0$ in the strong-coupling regime 
works to stem growth of the 
$u_{\mbox{\tiny {\rm BP}}}(\Psi_{0})$. 
This is a noticeable difference from 
the case of the weak-coupling regime 
(\ref{eq:lower-estimate-bipolaron-radius}).

We can show the above result 
in the same way as we did in Sec.\ref{sec:SLWCR}. 
Adopting Lieb's idea\cite{lieb} into our argument 
lead us the inequality, 
\begin{eqnarray*}
0 \ge 
-\,\alpha_{\theta}^{2}u_{\mbox{\tiny {\rm BP}}}(\Psi_{0}) 
-\,\sqrt{2}\alpha_{\theta}^{2} 
+ \frac{\sqrt{2}\,\alpha_{\theta}^{2}}{\theta(1-\eta)}\, 
\frac{|x_{j}|}{|x_{1}-x_{2}|}
- (4 + 3\theta)\sum_{k}|V_{k}|^{2}u_{\mbox{\tiny {\rm BP}}}(\Psi_{0}). 
\end{eqnarray*}
Here we used the inequalities $E_{\mbox{\tiny {\rm BP}}} \le 0$ 
and $p_{j'}^{2}/2 - \sqrt{2}/|x_{j'}| \ge -1$ 
for $j' \ne j$. 
Since $|x_{1}-x_{2}| \le |x_{1}| + |x_{2}|$, 
we eventually obtain 
\begin{eqnarray*}
0 \ge 
- 
\left\{ 2\alpha_{\theta}^{2} 
+ \left( 2 + \frac{3\theta}{2}\right)
E_{\mathrm {w}}(\alpha) 
\right\} u_{\mbox{\tiny {\rm BP}}}(\Psi_{0}) 
+ 
\sqrt{2}\,\alpha_{\theta}^{2}
\left( 
\frac{1}{\theta(1-\eta)}
- 2\right), 
\end{eqnarray*}
which implies the inequality (\ref{eq:SLSCR-result1}).

\section{Positive Binding Energy}
\label{sec:nonnegative-binding}

Combining the inequalities (\ref{theorem:BE1-2}) 
and (\ref{eq:upper-bound-to-gse}), 
we obtain a sufficient condition 
for the binding energy being positive. 
Namely, {\it if $c_{*}$, $\theta$, and $\eta$ satisfy 
$c_{\mbox{\tiny {\rm BP}}} > 2c_{\mbox{\tiny {\rm SP}}}$, 
then the binding energy is positive, i.e.,}
\begin{eqnarray}
E_{\mbox{\tiny {\rm BP}}} 
< 2E_{\mbox{\tiny {\rm SP}}}
\label{eq:non-negative-binding-energy}
\end{eqnarray}
{\it for sufficiently large $\theta > 0$ 
with the condition (\ref{eq:katei}).} 
Here, remember $c_{\mbox{\tiny {\rm SP}}} 
= 0.108513\cdots$. 
Then, we note that 
$0 < c_{\mbox{\tiny {\rm BP}}} \le 0.4$ 
and $\lim_{c_{*}\to 1, \theta\to\infty} = 0.4$ 
under the condition (\ref{eq:katei}). 
Moreover, the condition $c_{\mbox{\tiny BP}} 
> 2c_{\mbox{\tiny SP}}$ is equivalent to 
\begin{eqnarray}
\frac{2}{5}\left( 
c_{*} - \frac{1}{\theta(1-\eta)}
\right)^{2} 
> 0.217024, 
\label{eq:SD}
\end{eqnarray}
which has the form with corrections in 
the inequality (12) of Ref.\onlinecite{SD}. 
Thus, the condition 
(\ref{eq:SD}) is almost equivalent to 
\begin{eqnarray}
1 - \frac{1}{0.2635\theta} > 
c_{*} - \frac{1}{0.2635\theta} > \eta.
\label{eq:SD2}
\end{eqnarray}
As Smondyrev and Devreese\cite{SD} pointed out, 
we must keep it in mind that the inequality 
(\ref{eq:SD}) is a sufficient condition 
for the positive binding energy 
(\ref{eq:non-negative-binding-energy}), 
not a necessary condition. 
If we take the effect of not only the leading 
terms of $c_{\mbox{\tiny {\rm SP}}}$ and 
$c_{\mbox{\tiny {\rm BP}}}$ but also their 
remainders into the condition (\ref{eq:SD2}), 
we should control $c_{*}$ so that 
$\eta$ and $\theta$ meet to the results by 
experiments.   

According to the recent result of 
study\cite{MS} of the Hamiltonians 
$\sum_{j}^{\Pi=0}
\widetilde{H}_{\mbox{\tiny {\rm SP}}}^{(j)}$ 
and $\widetilde{H}_{\mbox{\tiny {\rm BP}}}^{\Pi=0}$, 
we might be able to choose $-\,\inf_{\varphi, 
\langle\!\langle\psi\rangle\!\rangle = 1}
{\mathcal E}_{\theta}(\varphi)$ as 
$c_{\mbox{\tiny BP}}$ (i.e., 
$-\, c_{\mbox{\tiny BP}} = \inf_{\varphi}
{\mathcal E}_{\theta}(\varphi)$) 
so that 
$2E_{\mbox{\tiny {\rm SP}}} - 
E_{\mbox{\tiny {\rm BP}}} 
= -\, (2c_{\mbox{\tiny {\rm SP}}} - 
c_{\mbox{\tiny {\rm BP}}})\alpha_{\theta}^{2} > 0$ 
for sufficiently large $\theta > 0$. 
We note that this type of equation 
expressing the binding energy 
is pointed out by Vinetskii and Giterman.\cite{VG,AM}

\begin{acknowledgements}
The author is grateful to T. Hattori, T. Miyao, 
G. L. Sewell, and H. Spohn for their useful comments. 
In particular, the author's eyes were reopened 
to the bipolaron problem through the joint works 
with H. Spohn. 
Thus, this work is dedicated to H. Spohn 
on the occasion of his $60$th birthday 
to express the author's thanks to him. 
The author greatly appreciates the two referees' 
thoughtful comments, which helped him 
to improve the results in his work. 
The author is also grateful to M. Sh. Birman, J. Dittrich, 
and P. Exner for giving some literature in Russian 
to the author.   
This work is supported by Japan 
Society for the Promotion of Science (JSPS), Grand-in-Aid for 
Scientific Research (C) 18540180. 
\end{acknowledgements}


\begin{thebibliography}{95}

\bibitem{shafroth}
M.~R.~Shafroth, 
Phys. Rev. {\bf 100}, 463 (1955).

\bibitem{anderson}
P.~W.~Anderson, 
Phys. Rev. Lett. {\bf 34}, 953 (1975).

\bibitem{AR}
A.~S.~Alexandrov and J.~Ranninger, 
Phys. Rev. B {\bf 23}, 1796 (1981); 
\textit{ibid}. {\bf 24}, 1164 (1981).

\bibitem{mott}
N.~F.~Mott,    
Nature (London) {\bf 327}, 185 (1987).

\bibitem{emin}
D.~Emin, 
Phys. Rev. Lett. {\bf 62}, 1544 (1989); 
D. Emin and M. S. Hillery,   
Phys. Rev. B {\bf 39}, 6575 (1989). 
 
\bibitem{FPZ}
H.~Fr\"{o}hlich, H.~Pelzer, and S.~Zienau, 
Philos. Mag. {\bf 41}, 221 (1950). 

\bibitem{froehlich}
H.~Fr\"{o}hlich, 
Adv. Phys. {\bf 3}, 325 (1954). 
 

\bibitem{froehlich0} 
H.~Fr\"{o}hlich, 
Phys. Rev. \textbf{79}, 845  (1950). 

\bibitem{BGIN}
F.~Bassani, M.~Geddo, G.~Iadonisi, and D.~Ninno, 
Phys. Rev. B {\bf 43}, 5296 (1991).

\bibitem{VSPD}
G.~Verbist, A.~Smondyrev, F.~M.~Peeters, and 
J.~T.~Devreese, 
Phys. Rev. B {\bf 45}, 5262 (1992).

\bibitem{VPD} 
G.~Verbist, F.~M.~Peeters, and 
J.~T.~Devreese, 
Phys. Rev. B {\bf 43}, 2712 (1991).

\bibitem{adamowski}
J.~Adamowski, 
Phys. Rev. B {\bf 39}, 3649 (1989).

\bibitem{tjablikov}
S.~V.~Tjablikov, 
Zh. Eksp. Teor. Fiz. {\bf 23}, 281 (1952).

\bibitem{YK}
J.~Yamashita and T.~Kurosawa, 
J. Phys. Chem. Solids. {\bf 5}, 34 (1958).

\bibitem{sewell}
G.~L.~Sewell, 
Philos. Mag. {\bf 3}, 1361 (1958).

\bibitem{holstein}
T.~Holstein, 
Ann. Phys. (N.Y.) {\bf 8}, 325 (1959); 
{\bf 8}, 343 (1959).

\bibitem{LF}
I.~G.~Lang and Yu.~A.~Firosov, 
Zh. Eksp. Teor. Fiz. {\bf 43}, 1843 (1962).

\bibitem{AK}
A.~S.~Alexandrov and P.~E.~Kornilovich, 
Phys. Rev. Lett. {\bf 82}, 807 (1999).

\bibitem{AM-small}
A.~S.~Alexandrov and N.~F.~Mott, 
Rep. Prog. Phys. {\bf 57}, 1197 (1994).

\bibitem{alexandrov}
A.~S.~Alexandrov, 
Phys. Rev. B {\bf 61}, 12315 (1999).

\bibitem{IL}
H.~Ibach and H.~L\"{u}th, 
\textit{Solid-State Physics} 
(Springer-Verlag, Berlin, 1990).

\bibitem{pekar}
S.~I.~Pekar, 
\textit{Issledovaniya po Elektronnoy Teorii Kristallov} 
(Gosud. Izd. Techn. 
--- Teor. Literatury, Moscow, 1951).

\bibitem{takada}
Y.~Takada, 
Phys. Rev. B {\bf 26}, 1223 (1982).

\bibitem{AM}
A.~S.~Alexandrov and S.~N.~Mott, 
\textit{Polarons \&  Bipolarons} 
(World Scientific, Singapore, 1995).

\bibitem{VG}
V.~L.~Vinetskii and M.~Sh.~Giterman, 
Z.~Eksp.~Teor.~Fiz. {\bf 33}, 730 (1957); 
V.~L.~Vinetskii, \textit{ibid}. {\bf 40}, 
1459 (1961).

\bibitem{HT}
H.~Hiramoto and Y.~Toyozawa,    
J. Phys. Soc. Jpn. {\bf 54}, 245 (1985).

\bibitem{feynman} 
R.~Feynman, 
Phys. Rev. \textbf{97}, 660  (1955). 

\bibitem{PD}
F.~Peeters and J.~T.~Devreese, 
J. Math. Phys. \textbf{21}, 2302  (1980). 

\bibitem{hirokawa-rims}
M.~Hirokawa, 
Publ. RIMS, {\bf 42}, 897  (2006).  

\bibitem{emin1}
D.~Emin, in 
\textit{Polarons and Bipolarons in High-$T_{c}$ Superconductors 
and Related Materials} edited by E.~K.~H. Salje, 
A.~S.~Alexandrov, and W.~Y.~Liang (Cambridge Univ. Press., 1995).

\bibitem{salje}
E.~K.~H.~Salje,    
in \textit{Polarons and 
Bipolarons in High-$T_{c}$ Superconductors 
and Related Materials} edited by E.~K.~H. Salje, 
A.~S.~Alexandrov, and W.~Y.~Liang (Cambridge Univ. Press., 1995).

\bibitem{LT}
E.~H.~Lieb and L.~E.~Thomas,    
Commun. Math. Phys. {\bf 183}, 511 (1997); 
\textit{ibid}. {\bf 188}, 499 (1997).

\bibitem{HHS}
M.~Hirokawa, F.~Hiroshima, and H.~Spohn,   
Adv. Math. {\bf 191}, 339  (2005).

\bibitem{lieb}
E.~H.~Lieb,    
Phys. Rev. Lett. {\bf 52}, 315 (1984); 
Phys. Rev. A {\bf 29}, 3018 (1984).

\bibitem{AM2}
N.~W.~Ashcroft and N.~D.~Mermin, 
\textit{Solid State Physics} 
(Brooks/Cole, Thomson Learning, Inc., 1976).

\bibitem{pines}
D.~Pines,    
\textit{Elementary Excitations in Solids} 
(Westview Press, 1999).

\bibitem{BB}
N.~N.~Bogolubov and N.~N.~Bogolubov, Jr., 
\textit{Polaron Theory: Model Problems} 
(Gordon and Breach Science Publishers, 2000).

\bibitem{kittel}
C.~Kittel, 
\textit{Quantum Theory of Solids} 
(John Wiley \& Sons, 1983).

\bibitem{HM}
M.~Hirokawa and T.~Miyao, 
(preprint, 2006).

\bibitem{LRU}
A.~De Luca, L.~M.~Ricciardi, and H.~Umezawa,    
Physica {\bf 40}, 61 (1968).

\bibitem{LLP}
T.~D.~Lee, F.~E.Low, and D.~Pines, 
Phys. Rev. {\bf 90}, 297  (1953). 

\bibitem{DV}
M.~D.~Donsker and S.~R.~S.~Varadhan,    
Commun. Pure Appl. Math. {\bf 36}, 595 (1983).

\bibitem{HHS2}
C.~Hainzl, M.~Hirokawa, and H.~Spohn,   
J. Funct. Anal. {\bf 220}, 424  (2005).

\bibitem{RS}
M.~Reed and B.~Simon, 
\textit{Methods of Modern Mathematical Physics I. 
Functional Analysis} 
(Academic Press, 1980).

\bibitem{miyake1}
S.~J.~Miyake,    
J. Phys. Soc. Jpn. {\bf 38}, 181 (1975); 
\textit{ibid}. {\bf 41}, 747 (1976).

\bibitem{lieb0}
E.~H.~Lieb,    
Stud. Appl. Math. {\bf 57}, 93 (1977).

\bibitem{spohn}
H.~Spohn,  
J. Phys. A {\bf 19}, 533 (1986); 
Phys. Rev. B {\bf 33}, 8906 (1986); 
Ann. Phys. {\bf 175}, 278 (1987); 
J. Phys. A {\bf 21}, 1199 (1988).

\bibitem{GL} 
B.~Gerlach and H.~L\"{o}wen, 
Rev. Mod. Phys. \textbf{63}, 63  (1991). 

\bibitem{MS}
T.~Miyao and H.~Spohn, 
\textit{} 
(preprint, 2006).

\bibitem{lakhno}
V.~D.~Lakhno, 
Phys. Rev. B {\bf 51}, 3512  (1995). 

\bibitem{SD}
M.~A.~Smondyrev and J.~T.~Devreese, 
Phys. Rev. B {\bf 53}, 11878  (1996). 

\bibitem{gross}
E.~P.~Gross,   
Ann. Phys. (N.Y.) {\bf 19}, 219  (1962); 
E.~Nelson,    
J. Math. Phys. {\bf 5}, 1190 (1964).

\bibitem{tomonaga}
S.~Tomonaga,    
Prog. Theor. Phys. {\bf 2}, 6 (1947).

\bibitem{LP1}
T.~D.~Lee and D.~Pines, 
Phys. Rev. {\bf 88}, 960  (1952); 
\textit{ibid}.  {\bf 92}, 883  (1953). 

\bibitem{kadanoff}
L.~P.~Kadanoff, 
Phys. Rev. {\bf 130}, 1364 (1963).

\bibitem{RS4}
M.~Reed and B.~Simon, 
\textit{Methods of Modern Mathematical Physics IV. 
Analysis of Operators} 
(Academic Press, 1978).



\end{thebibliography}
\end{document}